\documentclass[fleqn,12pt]{wlscirep}

\usepackage[utf8]{inputenc}
\usepackage[T1]{fontenc}
\DeclareUnicodeCharacter{00B5}{\textmu}
\DeclareUnicodeCharacter{00B0}{\ensuremath{{}^{\circ}}}
\usepackage{color,soul}
\usepackage{upgreek}
\usepackage{setspace}
\usepackage{amsmath}
\usepackage{mathtools}
\usepackage{amssymb}
\usepackage{bm}
\usepackage{siunitx}
\usepackage{etoolbox}
\usepackage{newtxtext}
\usepackage{newtxmath}
\usepackage{array}  
\usepackage{xcolor} 
\usepackage{geometry}  
\usepackage{pdflscape}  
\usepackage{longtable}  
\usepackage{longtable,array,xcolor,colortbl,booktabs}
\usepackage{float}

\makeatletter
\titleformat{\section}
  {\color{color1}\Large\rmfamily\bfseries}
  {\thesection}
  {0.5em}
  {#1}
  []
\titleformat{name=\section,numberless}
  {\color{color1}\Large\rmfamily\bfseries}
  {}
  {0em}
  {#1}
  []
\titleformat{\subsection}
  {\large\rmfamily\bfseries}
  {\thesubsection}
  {0.5em}
  {#1}
  []
\titleformat{name=\subsection,numberless}
  {\large\rmfamily\bfseries}
  {}
  {0em}
  {#1}
  []
\titleformat{\subsubsection}
  {\normalsize\rmfamily\bfseries\itshape}
  {\thesubsubsection}
  {0.5em}
  {#1}
  []
\titleformat{\paragraph}[runin]
  {\normalsize\rmfamily\bfseries}
  {}
  {0em}
  {#1}
\renewcommand{\@maketitle}{%
{%
\thispagestyle{plain}%
\vskip-36pt%
{\centering\rmfamily\bfseries\fontsize{20}{25}\selectfont \@title\par}%
\vskip10pt
{\centering\rmfamily\fontsize{14}{18}\selectfont  \@author\par}
\vskip18pt%
{\noindent{\color{color1}\Large\rmfamily\textbf{Abstract}}\par}%
\vskip8pt
{\noindent\rmfamily\normalsize\bfseries\theabstract\par}%
\vskip25pt%
}%
}%
\makeatother
\title{\centering \textcolor{black}{Multi-channel Optical Vision Model}}

\author[1]{Ali Momeni}
\author[1]{Guillaume Noetinger}
\author[1]{Tim Tuuva}
\author[1,*]{Romain Fleury}

\affil[1]{Laboratory of Wave Engineering, School of Electrical Engineering, Swiss Federal Institute of Technology in Lausanne (EPFL), Lausanne, Switzerland}

\affil[*]{E-mail: romain.fleury@epfl.ch}

\begin{abstract}
Spatial multiplexing is one of the natural strengths of optics, yet in optical neural networks it is often used mainly as parallel throughput. Here we show that spatial multiplexing in an optical neural network can be used not only to process multiple inputs in parallel, but also to define a trainable representational coordinate of the model. In three implemented scenarios, parallel-input processing, class-code readout and channel-mixed feature interaction, spatial channels act as independent learners, structured code dimensions and interacting feature groups. The programmable free-space optical processor is trained through an online physical-forward/surrogate-backward scheme, where measured optical outputs define the forward pass while a differentiable surrogate estimates gradients and is continually fine-tuned during training from newly acquired optical data. We demonstrate these channel roles in image-classification and regression tasks using multi-layer architectures with more than one million trainable optical phase parameters. We further implement a hybrid optical-electronic vision-language model, in which the optical neural network provides visual tokens to a digital transformer decoder for controlled image-captioning tasks. These results establish spatially multiplexed optical channels as a programmable feature and readout space for hybrid optical vision models.
\end{abstract}

\begin{document}

\maketitle

\section*{Introduction}

Modern vision models, including convolutional neural networks and more recent vision transformers, rely on dense numerical operations, repeated memory access and increasingly large training and inference pipelines \cite{krizhevsky2012imagenet,he2016deep,dosovitskiy2021image}. This computational load places growing pressure on conventional electronic hardware, particularly when visual or multimodal inputs must be encoded, moved through memory and transformed repeatedly before a final decision is made \cite{lecun2015deep,sebastian2020memory,wetzstein2020inference}. Optical and photonic systems offer a complementary computational substrate in which diffraction, interference, modulation and detection act directly on high-dimensional spatial fields. This makes optics naturally suited to visual processing: an image is already a spatial field, and optical propagation can transform many spatial degrees of freedom in parallel. The central question is therefore not only whether optics can accelerate isolated linear operations, but whether programmable optical systems can be trained as visual front ends whose physical degrees of freedom participate in representation learning.

Optical neural networks have advanced along several directions. Diffractive and free-space optical processors have shown that cascaded optical layers can implement learned spatial transformations and task-specific inference \cite{lin2018all,liu2022programmable,mengu2020analysis,isil2024denoising}. Integrated photonic systems have demonstrated optical acceleration of matrix multiplication, convolution, and neural-network primitives, with high bandwidth and compact implementations \cite{shen2017deep,feldmann2021parallel,xu2021tops,hua2025integrated,ahmed2025universal}. Optical and optoelectronic vision processors further suggest that computation can be embedded into sensing or analog front ends \cite{chen2023allanalog,wang2023image}. At the training level, physical neural networks and forward-mode optical training have emphasized that measured hardware responses can be included directly in optimization, reducing the mismatch between idealized simulation and experimental behavior \cite{wright2022deep,xue2024fully}. Related work has also examined backpropagation-free physical learning, broader training strategies for physical neural networks, programmable-metasurface expressivity, and nonlinear time-Floquet wave learning \cite{momeni2023backpropagation,momeni2025training,hammami2026expressivity,momeni2022electromagnetic}. Recent work on nonlinear encoding, optical transformers, large-scale photonic accelerators, and optical generative models points toward more complex optical AI workflows \cite{xia2024nonlinear,yildirim2024nonlinear,wanjura2024fully,anderson2023optical,xu2024taichi,chen2025optical}. However, spatial multiplexing in many demonstrations remains closer to parallel throughput, independent sensing, or task-specific readout than to a trainable representational axis.

The key challenge is therefore not only to build larger optical processors, but to define optical architectures in which multiplexed channels carry structured computational meaning. This reveals a fundamental challenge: multi-channel optical representation learning must be realized directly on imperfect physical hardware, where experimental constraints and system nonidealities play a central role, rather than in idealized simulations alone. Training such systems requires end-to-end optimization through the physical hardware while accounting for channel-dependent aberrations, finite modulator and detector precision, optical cross-talk, measurement noise, fluctuations in signal-to-noise ratio, and laser drift. It also requires a differentiable interface that scales across channels. Training an independent surrogate for each channel would discard the shared structure of optical propagation and would scale poorly with channel number. A single channel-conditioned surrogate offers a practical way to share information across channels while retaining channel-specific corrections, and online fine-tuning can update this surrogate from measured input-output pairs collected during training.

Here we show that spatial multiplexing in an optical neural network can be used not only to process multiple inputs in parallel, but also to define a trainable representational coordinate of the model. We experimentally implement a multi-layer and multi-channel optical vision architecture trained through an online physical-forward/surrogate-backward scheme, in which measured optical outputs define the forward pass and a differentiable surrogate supplies gradients while being continually updated from optical data. The optical layers also use an input-dependent structural nonlinearity at the {spatial light modulators (SLM)} plane, where the channel input is embedded into the displayed physical phase modulation. We investigate three scenarios of channel-based computation: parallel-input processing, where spatial channels act as independent optical learners; class-code readout, where channels form structured code dimensions for classification; and channel-mixed feature interaction, where measured channel outputs are coupled between optical layers. We demonstrate these roles on benchmark vision tasks including handwritten-digit classification (MNIST), fashion-item classification (Fashion-MNIST), and facial keypoint regression, using multi-layer optical architectures with more than one million trainable phase parameters. We further extend the framework to a hybrid optical-electronic vision-language model, in which the optical neural network provides visual tokens to a digital transformer decoder for controlled image captioning. Together, these results establish spatially multiplexed optical channels as a programmable feature and readout space for hybrid optical vision models.
\begin{figure}[H]
    \centering
    \includegraphics[width=1\textwidth]{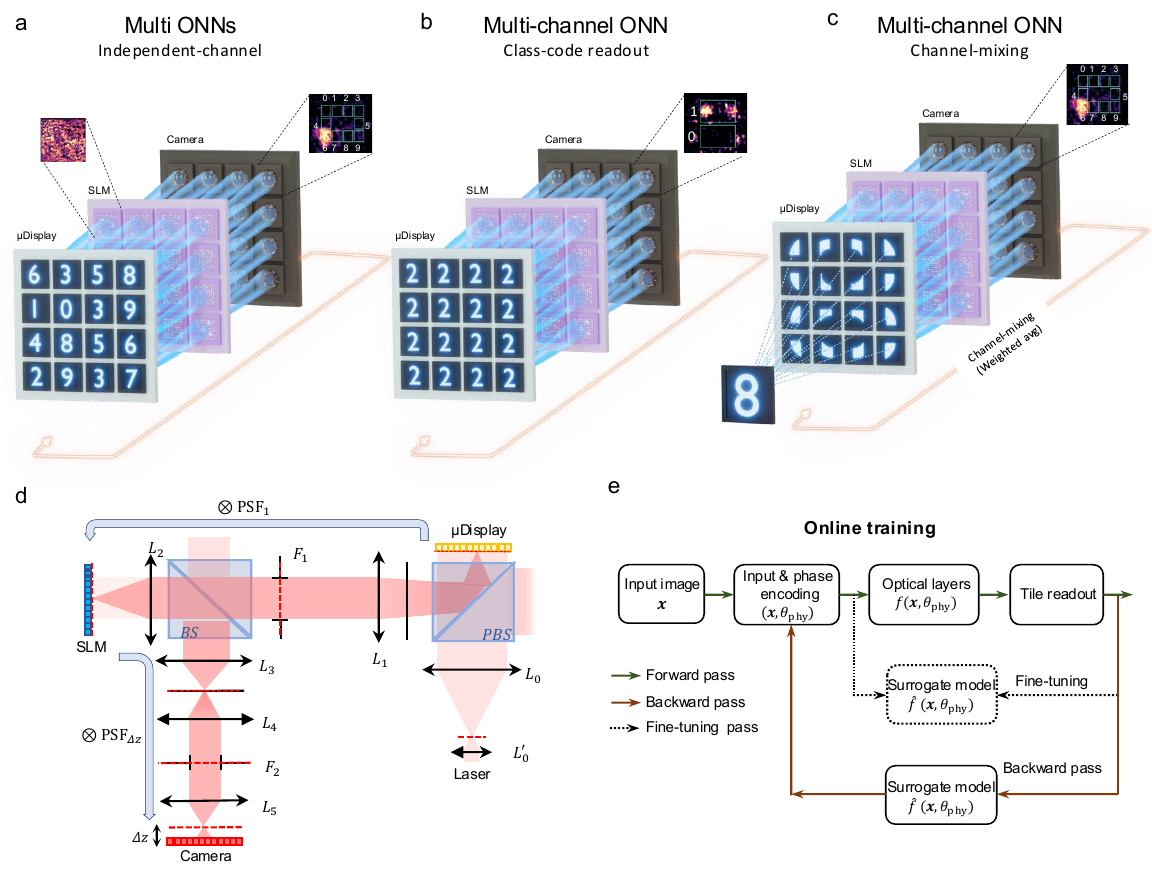}
    \caption{\textcolor{black}{\textbf{Multi-channel optical vision model and experimental training workflow.} \textbf{a,} Independent-channel multi-ONN readout. Different input samples are assigned to spatial channels on the microdisplay, encoded by channel-specific phase masks on the SLM, propagated through the free-space optical stack, and measured as separate camera-output tiles. Each channel can therefore act as an independent optical learner. \textbf{b,} Class-code readout. The same input image is repeated across optical channels, and the measured channel outputs are converted into bit-like readout margins that are decoded against a class code. \textbf{c,} Channel-mixing mode. A single input image is encoded across multiple optical channels, and the measured outputs are coupled by learned electronic \(16\!\rightarrow\!16\) channel mixers before being re-encoded at the next optical layer. This coupling lets distributed channel features interact while the optical transformations remain spatially multiplexed. \textbf{d,} Free-space optical implementation: {an expanded laser beam is directed onto a liquid cristal amplitude modulator (\textit{the microdisplay}) displaying the input data. The resulting field is sent to a phase-only SLM to apply some learnable weights in a conjugate plane. The modulated beam is then filtered and collected by a defocused camera. This entire optical path }defines the measured optical forward pass. {It is equivalent to the propagation in a slab with tunable transmission.} \textbf{e,} Online physical-forward/surrogate-backward training. The optical system produces measured outputs during the forward pass, a single differentiable surrogate supplies gradients for backpropagation, and buffered optical input-output pairs are used to fine-tune the surrogate during training.}}
    \label{fig:fig1}
\end{figure}
\section*{Results}

\subsection*{Architecture and experimental setup}

We first establish the architecture and training workflow of the proposed multi-layer, multi-channel optical vision model (Fig.~\ref{fig:fig1}). The system is formulated for \(N\) spatially multiplexed optical channels arranged as tiled fields on the input and output planes. In a general form, the field in channel \(c\), with \(c=1,\ldots,N\), at optical layer \(l+1\) can be written as
\begin{equation}
\label{eq:channel_forward}
\mathbf{x}_{c}^{(l+1)}
=
f\!\left(
\mathbf{x}_{c}^{(l)},\boldsymbol{\theta}_{\mathrm{phys},c}^{(l)}
\right),
\end{equation}
where \(\mathbf{x}_{c}^{(l)}\) denotes the spatial intensity pattern encoded in channel \(c\) at layer \(l\), \(\boldsymbol{\theta}_{\mathrm{phys},c}^{(l)}\) denotes the displayed trainable physical phase profile for that channel and layer, and \(f\) denotes the measured physical transformation including intensity encoding, phase modulation, free-space propagation, optical detection and camera readout. The demonstrations illustrated in Fig.~\ref{fig:fig1}a--c differ in how the channel inputs are defined and how the channel outputs are combined. In the parallel-input (independent-channel) scenario, different images are assigned to different channels, \(\mathbf{x}_{c}^{(0)}=\mathbf{I}_{c}\), and each channel acts as an independent optical learner (Fig.~\ref{fig:fig1}a). The optical outputs are also read out independently, allowing each channel to produce its own prediction without requiring information from the other channels. In the MNIST demonstration shown in Fig.~\ref{fig:fig1}a, the camera readout associated with each channel is divided into ten spatial detector tiles, with each tile corresponding to one digit class. The integrated intensity measured within these tiles forms the channel-specific output vector, and the predicted class is obtained from the tile receiving the highest optical energy. Because the channels operate in parallel while sharing the same optical hardware, multiple images can be processed simultaneously, each with its own dedicated optical path, trainable phase masks and decision readout. In the class-code scenario, the same image is replicated across all channels, \(\mathbf{x}_{c}^{(0)}=\mathbf{I}\), and each channel is read out through two spatial tiles that define a binary response; across channels, these responses form a structured multi-bit class code (Fig.~\ref{fig:fig1}b). In the channel-mixed scenario, a single image is encoded into channel-specific inputs, \(\mathbf{x}_{c}^{(0)}=\mathcal{E}_{c}(\mathbf{I})\), and the measured outputs are coupled by learned electronic \(16\!\rightarrow\!16\) channel mixers between optical layers (Fig.~\ref{fig:fig1}c). This last case is central to the representational use of multiplexing: information is not only read out from parallel channels, but exchanged across channels during the layered computation. The task-specific input encodings and optical readout definitions are summarized in Supplementary Section~\ref{sec:supp_task_details} and Supplementary Table~\ref{tab:task_interface}.

The experimental platform is a programmable free-space optical processor (Fig.~\ref{fig:fig1}d) {made by the conjugation of an amplitude and a phase spatial light modulators (respectively named the \textit{microdisplay} and \textit{SLM} hereafter) sent to a defocused camera mimicking propagation through a diffractive layer with a tunable transmission (see Methods). Hence, successive propagations through the setup are equivalent to the propagation in a layered complex media with non-linearities, here the camera readout.}  In the experiments reported here, \(N=16\) channels are arranged as a \(4\times4\) tiled optical field, {they are aligned thanks to a semi-automatic procedure in order to have the same spatial response \cite{noetinger2026tutorial}}. A detailed optical layout is provided in Supplementary Fig.~\ref{fig:DetailedSetup}. A laser-illuminated microdisplay encodes the channel intensity patterns, relay optics map the modulated light to the spatial light modulator, and channel-dependent trainable phase masks are applied before propagation to the camera. For a single channel {$c$} and optical layer {$l$}, the measured update can be written as an effective intensity-to-intensity transformation,
\begin{equation}
\label{eq:physical_channel_forward}
\mathbf{x}_{c}^{(l+1)}
=
\mathcal{D}\!\left[
\left|
h_{\Delta z} \ast
\left(
\exp\!\left(\mathrm{i}\,2\pi\boldsymbol{\theta}_{\mathrm{phys},c}^{(l)}\right)
\odot
\left[
h_{1} \ast \mathcal{A}\!\left(\mathbf{x}_{c}^{(l)}\right)
\right]
\right)
\right|^{2}
\right],
\end{equation}
where \(\mathcal{A}\) denotes the display and illumination encoding, \(h_{1}\) and \(h_{\Delta z}\) are effective point-spread functions for the relay and propagation sections shown in Fig.~\ref{fig:fig1}d, \(\odot\) denotes pixelwise phase modulation on the spatial light modulator, {$*$ is the 2D convolution operator}, and \(\mathcal{D}\) includes camera readout, cropping, normalization and preparation of the measured image for the next layer. This expression summarizes the optical computation performed by modulation, free-space propagation and intensity detection; during training, however, the forward value is the measured camera response of the physical system.

\begin{figure}[H]
    \centering
    \includegraphics[width=0.90\textwidth]{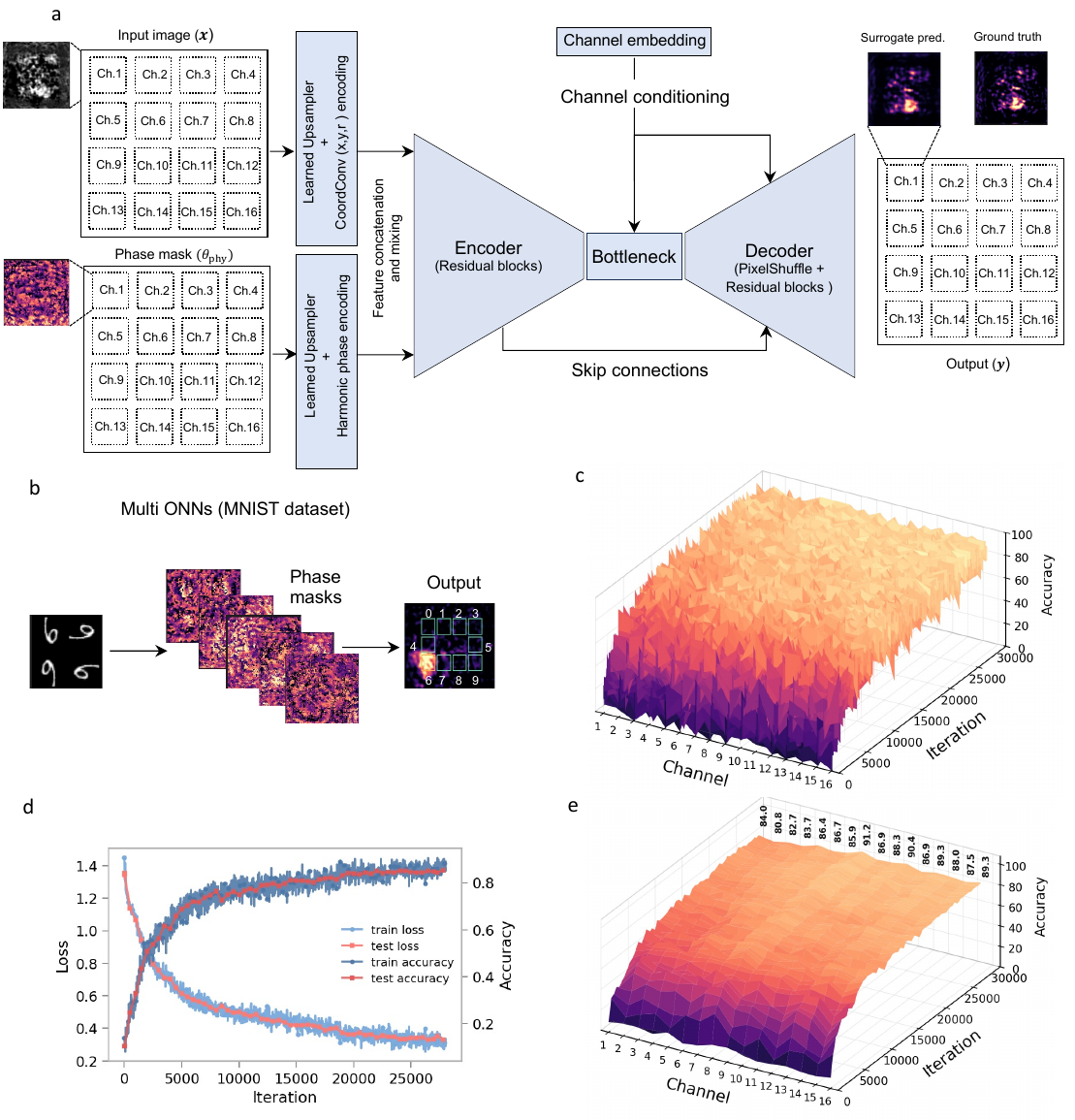}
    \caption{\textcolor{black}{\textbf{Channel-conditioned surrogate model and MNIST multi-ONN training.} \textbf{a,} Surrogate architecture used to approximate the optical response across all 16 channels. The model encodes the input field and phase mask using learned upsampling, coordinate features, and multi-harmonic phase features, then predicts the camera-plane output with an encoder-decoder network, skip connections, and channel conditioning. Example surrogate predictions are compared with measured or reference optical outputs. \textbf{b,} Example five-layer MNIST multi-ONN inference, showing the tiled digit input, trainable phase masks, and camera-plane class readout. \textbf{c,} Channel-wise training and validation accuracy surfaces over iteration for the 16 optical channels. \textbf{d,} Training and validation loss/accuracy curves aggregated across the 16 channel-wise classifiers, showing that all channels are trained jointly in one multi-channel ONN. \textbf{e,} Channel-wise validation performance map summarizing how the distribution of channel accuracies stabilizes during optimization.}}
    \label{fig:fig2}
\end{figure}

Training is performed with an online physical-forward/surrogate-backward scheme (Fig.~\ref{fig:fig1}e). In the forward pass, the encoded input and displayed optical phase profile are sent through the physical processor, so that the layer output is the measured camera response, \(\mathbf{y}_{\mathrm{phys}}=f\!\left(\mathbf{x},\boldsymbol{\theta}_{\mathrm{phys}}\right)\), rather than the output of a numerical optical model. For ONN training, a physics-informed differentiable surrogate \(\hat{f}\!\left(\mathbf{x},\boldsymbol{\theta}_{\mathrm{phys}}\right)\) (Fig.~\ref{fig:fig2}a) is evaluated on the same encoded input and phase profile and is used only to supply gradients during the backward pass. Thus, the loss is computed from the measured optical readout, while the parameter update is obtained through the surrogate approximation. This workflow enables multi-layer optical architectures with more than one million trainable phase parameters to be optimized directly through measured optical forward passes. The surrogate architecture and pre-training objective are detailed in Supplementary Section~\ref{sec:supp_surrogate_model}, Supplementary Table~\ref{tab:surrogate_architecture} and Supplementary Eq.~\ref{eq:surrogate_loss}.

The surrogate uses the intensity-to-intensity model in Eq.~\ref{eq:physical_channel_forward} as an approximate optical model: measured point-spread-function (PSF) kernels corresponding to the relay and propagation terms \(h_{1}\) and \(h_{\Delta z}\) provide propagation priors, and trainable encoder-decoder modules refine this approximation. As illustrated in Fig.~\ref{fig:fig2}a, tiled input-image and phase-mask channels are first mapped through learned upsampling and coordinate or harmonic phase encodings, concatenated into a shared feature representation, and passed through a U-Net-style encoder, bottleneck and decoder with residual blocks, skip connections and channel embeddings. This refinement path {learns} residual aberrations, {parasitic reflections and corresponding interferences,} channel-dependent deviations and readout differences; the representative surrogate output shown in Fig.~\ref{fig:fig2}a has a generated-image MSE of about \(5\times10^{-4}\) relative to the ground-truth optical output. Channel-resolved comparisons between measured and predicted optical outputs are shown in Supplementary Figs.~\ref{fig:surrogate_channel_examples_a} and~\ref{fig:surrogate_channel_examples_b}. A single surrogate is shared across the implemented channel array. During training, newly measured optical pairs \(\left(\mathbf{x},\boldsymbol{\theta}_{\mathrm{phys}},\mathbf{y}_{\mathrm{phys}}\right)\) are stored in a buffer and used to fine-tune \(\hat{f}\) online, allowing the differentiable model to track the physical system as the optical parameters evolve.

\subsection*{Training ONN for vision tasks}

We first train the multi-channel ONN in the independent-channel mode using MNIST classification (Fig.~\ref{fig:fig2}b--e). Different digit samples occupy different spatial channels, propagate through channel-specific trainable phase masks, and are read out from camera-plane detector tiles. The integrated intensity in these detector regions forms a digit-score vector for each channel, so each channel produces its own class prediction while the optical hardware is used in parallel. The input/readout definitions, optimization objectives and validation metrics for the five demonstrations are summarized in Supplementary Tables~\ref{tab:task_interface} and~\ref{tab:optimization_protocol}.

In this demonstration, the independent-channel model is implemented as a five-layer optical stack with 16 spatial channels and 1,003,520 trainable phase values, providing a direct test of whether many physical phase degrees of freedom can be coordinated across a shared multiplexed field. The corresponding trained phase masks are shown in Supplementary Fig.~\ref{fig:phase_masks_independent_mnist}. To increase the effective nonlinear interaction within this otherwise linearly propagating optical system, we also introduce a structural nonlinearity at the SLM plane. Here, structural nonlinearity denotes an input-dependent physical parameterization: the input pattern is embedded into the displayed SLM modulation, so that the optical operator depends jointly on the data and on the trainable phase profile rather than on the phase profile alone. This follows recent physical and optical neural-network schemes in which input-dependent scattering potentials, repeated data encodings, structural nonlinearities, or nonlinear optical encoders generate nonlinear functions using linear wave propagation \cite{momeni2023backpropagation,hammami2026expressivity,wanjura2024fully,yildirim2024nonlinear,xia2024nonlinear}. Consistent with Eq.~\ref{eq:channel_forward}, for channel \(c\) at layer \(l\), we write the displayed physical phase as \(\boldsymbol{\theta}_{\mathrm{phys},c}^{(l)} = \mathbf{T}_{c}^{(l)} \odot \mathbf{x}_{c}^{(l)} + \boldsymbol{\theta}_{\mathrm{train},c}^{(l)}\), where \(\mathbf{x}_{c}^{(l)}\) is the encoded input to that channel and layer, \(\boldsymbol{\theta}_{\mathrm{train},c}^{(l)}\) is the trainable phase profile, \(\mathbf{T}_{c}^{(l)}\) controls the strength and spatial weighting of the input-dependent term, and \(\odot\) denotes pixelwise multiplication. This data-dependent modulation creates multiplicative interactions between the input, trainable phase, and diffractive propagation, expanding the optical feature space without requiring a separate nonlinear optical material. The channel-wise accuracy surface in Fig.~\ref{fig:fig2}c shows that the 16 channels improve over training with broadly similar trajectories, although some channel-dependent variation remains. The training and validation curves in Fig.~\ref{fig:fig2}d report the loss and accuracy aggregated across the channel-wise classifiers, showing that all channels are trained together within one multi-channel ONN rather than optimized as separate networks. The map in Fig.~\ref{fig:fig2}e summarizes this evolution across channels; the mean of the final channel-wise validation accuracies is 86.9\%, with individual channels ranging from 80.8\% to 91.2\%. These results show that spatially multiplexed channels can be trained as independent optical learners, while the remaining spread across channels indicates residual channel imbalance in the measured multi-channel system.

\begin{figure}[H]
    \centering
    \includegraphics[width=0.90\textwidth]{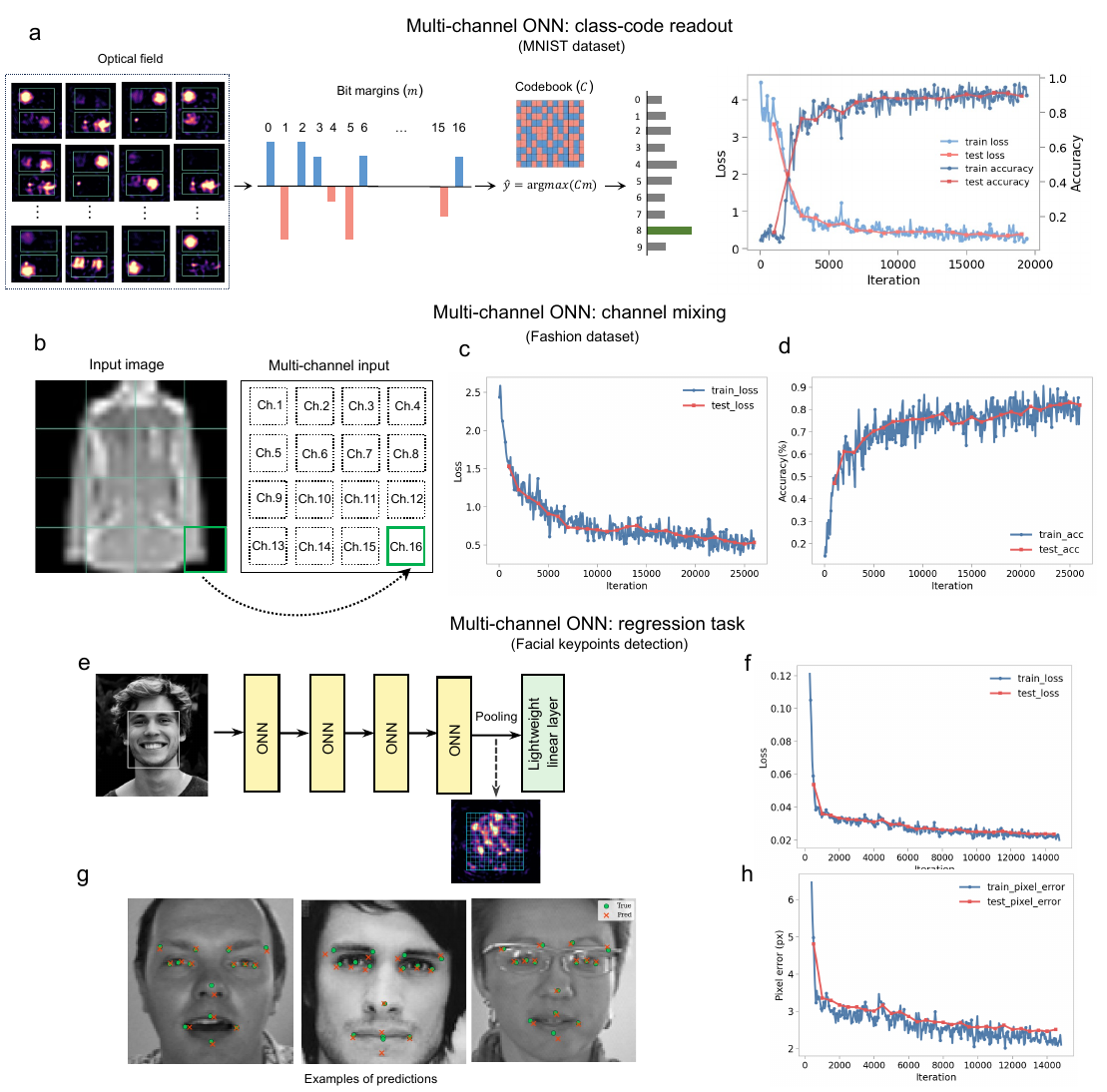}
    \caption{\textcolor{black}{\textbf{Class-code readout, channel mixing, and facial-keypoint regression in multi-channel ONNs.} \textbf{a,} Class-code readout on MNIST. The tiled camera output is converted into channel-wise bit margins and decoded against a fixed \(10\times16\) class-code book; class scores are computed from code-matched projections, and the predicted class is selected by the largest score. \textbf{b,} Channel input encoding for Fashion-MNIST mixing. A Fashion-MNIST image is mapped onto the 16 optical channels before learned electronic inter-layer mixing. \textbf{c,} Training and validation loss for the five-layer Fashion-MNIST channel-mixing run. \textbf{d,} Training and validation accuracy for the same channel-mixing run, showing the fused classification performance. \textbf{e,} Facial-keypoint regression architecture. A face image is processed by the multi-channel optical stack, and the measured optical features are pooled and mapped by per-channel electronic linear heads with learned channel fusion to 30 normalized landmark coordinates. \textbf{f,} Training and validation regression loss for facial-keypoint prediction. \textbf{g,} Representative held-out facial-keypoint predictions, with ground-truth landmarks in green and predicted landmarks in red. \textbf{h,} Training and validation coordinate-RMSE curves for facial-keypoint prediction, computed from normalized-coordinate MSE after conversion to the 96-pixel image scale.}}
    \label{fig:fig3}
\end{figure}

We next implement a class-code readout, in which the channel axis is used as a structured output space rather than as a set of independent classifiers (Fig.~\ref{fig:fig3}a). In this mode, the same input image is replicated across all optical channels, and each channel is read out through two spatial detector tiles corresponding to a binary response. The resulting channel-wise bit margins are decoded using a fixed codebook with 16 binary channels for 10 classes. The codebook is balanced: each class codeword contains eight \(+1\) and eight \(-1\) entries, and each channel assigns five classes to the positive bit and five classes to the negative bit. The minimum pairwise Hamming distance between class codewords is 8, with a mean distance of 8.89; specifically, 25 class pairs have distance 8 and 20 class pairs have distance 10, so every pair of classes differs in at least half of the 16 channels. This design turns the optical channel axis into a distributed class representation, where decisions are formed from the collective response of many channels rather than from a single readout. As a result, weak or noisy channels can be partially compensated by the remaining channel responses, and the large codeword separation provides robustness to channel imbalance and bit-level readout errors. Without inter-layer channel mixing or a trainable electronic feature mixer, the class-code optical model reaches 91.50\% held-out validation accuracy, showing that spatially multiplexed channels can serve as a robust structured readout dimension even when the optical channels remain independent until final codebook decoding. The corresponding trained phase masks are shown in Supplementary Fig.~\ref{fig:phase_masks_class_code}.

Channel mixing provides a direct test of whether spatially multiplexed optical outputs can become an interacting feature representation rather than a collection of independent readouts. In the Fashion-MNIST demonstration, each input image is encoded across the 16 optical channels through a learned electronic input adapter (Fig.~\ref{fig:fig3}b). After each optical layer, learned electronic \(16\!\rightarrow\!16\) channel mixers combine the measured outputs before they are re-encoded into the next optical layer. This hybrid coupling is not all-optical mixing; rather, it provides a controlled mechanism for exchanging information between optical channels while preserving the spatially multiplexed optical computation. The advantage is that distributed optical features are no longer forced to remain isolated: evidence across channels can be integrated over successive layers, weak or noisy responses can be compensated by the remaining channels, and the channel axis can act as an interacting feature dimension. In a five-layer optical architecture, this channel-mixed model reaches 83.17\% held-out validation accuracy, with a best validation loss of 0.5134 (Fig.~\ref{fig:fig3}c,d). These results show that learned electronic inter-channel mixing can turn spatial multiplexing from parallel optical processing into a trainable distributed representation for visual inference. The corresponding trained phase masks are shown in Supplementary Fig.~\ref{fig:phase_masks_fashion_mixing}.

We then extend the multi-channel optical vision model from classification to continuous visual regression (Fig.~\ref{fig:fig3}e--h; trained phase masks in Supplementary Fig.~\ref{fig:phase_masks_face_regression}). The task is facial-keypoint regression: given a face image, the model predicts the two-dimensional locations of 15 facial landmarks, corresponding to 30 normalized output coordinates. The input face is partitioned across the 16 optical channels and processed through the multi-layer optical stack with channel mixing (Fig.~\ref{fig:fig3}e). The measured optical outputs are pooled and mapped to landmark coordinates by per-channel electronic linear heads, whose predictions are combined through learned channel-fusion weights. During training, the regression objective combines coordinate error with an output-mask penalty, and the reported coordinate RMSE is computed from the normalized-coordinate MSE as \(\sqrt{\mathrm{MSE}}\times96\) pixels. The five-layer optical model reaches a best validation MSE of \(6.57\times10^{-4}\), corresponding to a coordinate RMSE of 2.461 pixels (Fig.~\ref{fig:fig3}f,h). Representative held-out predictions localize landmarks around the eyes, nose, mouth and face contour (Fig.~\ref{fig:fig3}g). These results show that the hybrid multi-channel optical front end can support continuous spatial readout, extending the channel-based optical representation beyond class decisions to coordinate-level visual inference.

\subsection*{Optical vision-language model}

\begin{figure}[H]
    \centering
    \includegraphics[width=0.90\textwidth]{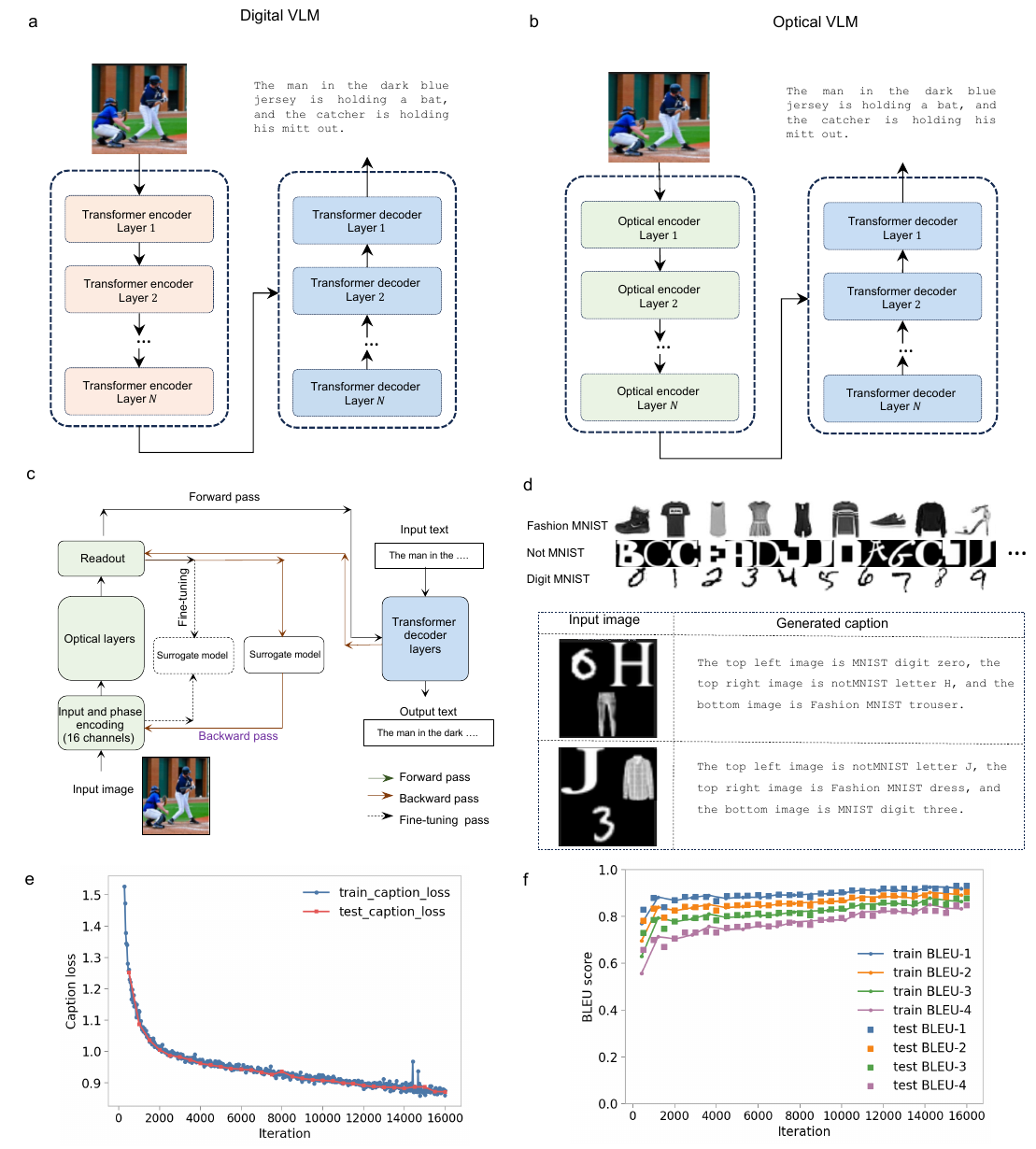}
    \caption{\textcolor{black}{\textbf{Optical vision-language model with a digital transformer decoder.} \textbf{a,} Digital encoder-decoder captioning baseline, shown for architectural comparison. \textbf{b,} Hybrid optical-digital vision-language architecture, in which the optical stack replaces the visual encoder/front end while the transformer decoder remains digital. \textbf{c,} Training workflow for the optical vision-language model. The image is encoded into optical channels, propagated through optical layers, and read out as visual tokens for the decoder; the backward pass uses a differentiable surrogate, with optional surrogate fine-tuning from measured optical pairs. \textbf{d,} Controlled captioning dataset, composed from MNIST, Fashion-MNIST, and NotMNIST image slots with captions describing the top-left, top-right, and bottom objects. \textbf{e,} Training and validation caption cross-entropy loss for the optical vision-language run. \textbf{f,} Training and validation BLEU-1 through BLEU-4 scores for generated captions on the controlled captioning task.}}
    \label{fig:fig4}
\end{figure}

Vision-language models commonly pair a visual encoder with a language decoder: the encoder converts an image into a sequence of visual tokens, and the decoder uses these tokens to generate text \cite{vaswani2017attention,dosovitskiy2021image,vinyals2015show,xu2015show,radford2021learning}. In a fully digital implementation, both the visual-token formation and the autoregressive text generation are performed electronically (Fig.~\ref{fig:fig4}a). Here we use the multi-channel optical neural network as the visual encoder/front end, while retaining a digital transformer decoder for sequence generation (Fig.~\ref{fig:fig4}b). In this hybrid architecture, optical propagation and trainable phase modulation produce image-conditioned visual tokens, which are then passed to the decoder in place of tokens produced by a digital transformer encoder. Thus, the optical system replaces the encoder-side visual computation of an encoder--decoder pipeline, while the language modelling, attention over text history and token generation remain digital. This provides a controlled starting point for optical vision-language processing: the optical network is not used merely as a classifier, but as a trainable visual-token generator for a downstream sequence model.

The training workflow follows the same physical-forward/surrogate-backward principle used in the classification and regression experiments (Fig.~\ref{fig:fig4}c). The input image is encoded into 16 optical channels, propagated through a six-layer optical stack and read out as a compact set of visual features for the transformer decoder. During the forward pass, these visual tokens are obtained from the measured optical response. During the backward pass, gradients with respect to the trainable optical phase parameters are supplied by the differentiable surrogate, with online fine-tuning from measured optical pairs. The decoder is trained to generate captions conditioned on the optical tokens and the previous text tokens. In this sense, the optical front end participates in the sequence-generation task through the visual representation it provides, while the text-generation mechanism itself remains electronic. The captioning task interface and optimization protocol are summarized in Supplementary Tables~\ref{tab:task_interface} and~\ref{tab:optimization_protocol}, and the trained six-layer optical phase masks are shown in Supplementary Fig.~\ref{fig:phase_masks_vlm}.

We evaluate this architecture on a controlled compositional captioning dataset designed to test whether the optical encoder can preserve object identity and spatial structure in a form usable by the decoder (Fig.~\ref{fig:fig4}d). Each image canvas is assembled from objects sampled from MNIST digits, Fashion-MNIST categories and NotMNIST letters, placed in fixed spatial slots such as the top-left, top-right and bottom regions. The target caption describes both the identity and position of the objects, for example by specifying which digit, letter or fashion item appears in each slot. This task is intentionally more structured than open-vocabulary natural-image captioning, but it requires the optical encoder to provide visual tokens that retain multi-object and positional information across the channel array.

The hybrid optical-electronic model learns stable caption generation on this controlled task (Fig.~\ref{fig:fig4}e,f). The validation caption loss decreases throughout training and reaches a final value of 0.8712. Caption quality is evaluated using BLEU scores, a standard sequence-generation metric that compares generated captions with reference captions through modified \(n\)-gram precision and a brevity penalty. BLEU-1 measures word-level agreement, whereas BLEU-2, BLEU-3 and BLEU-4 progressively test longer phrase overlap and therefore provide stricter measures of sentence structure. In our controlled captioning task, high BLEU scores indicate that the decoder recovers both the object identities and the compositional caption template from the optical visual tokens. The final validation BLEU-1 through BLEU-4 scores are 0.9299, 0.9045, 0.8774 and 0.8479, respectively. More broadly, this result extends the role of the multi-channel optical neural network beyond fixed readout tasks: the same spatially multiplexed optical front end can supply task-relevant visual representations to a sequence model. Although the present demonstration uses a controlled captioning dataset and a digital decoder, it establishes a first step toward replacing digital visual encoders with trainable optical encoders in hybrid vision-language systems.

\section*{Discussion}

This work shows that spatial multiplexing in optical neural networks can be used for more than parallel throughput. Across the demonstrations, the channel axis is progressively promoted from a hardware layout into a computational coordinate of the model. In the independent-channel experiment, multiplexing allows several optical learners to be trained within the same physical processor. In the class-code readout, the same channel array becomes a distributed output space, where a prediction is formed collectively from many binary optical responses rather than from a single detector region. This representation introduces redundancy: weak, noisy or imbalanced channels do not necessarily determine the final decision, because the codeword separation allows other channels to compensate. In the channel-mixing experiment, multiplexed optical outputs are no longer only read out in parallel, but are coupled between layers, allowing information distributed across channel encodings to interact during the computation. Together, these regimes demonstrate that the channel dimension can carry structured computational meaning.

The extension from classification to facial-keypoint regression and controlled caption generation further broadens this view. The regression task shows that the multi-channel optical front end can support continuous spatial outputs, not only discrete class decisions. The vision-language experiment then tests a more general role for the optical network: rather than acting as a classifier, it produces visual tokens for a downstream digital transformer decoder. This does not make the system a fully optical vision-language model, but it establishes an important architectural possibility. The optical processor can replace the visual-encoder side of an encoder--decoder pipeline, while the language decoder remains electronic. In this hybrid form, the optical system contributes trainable visual representations that are useful for sequence generation, pointing toward optical front ends that interface naturally with modern token-based machine-learning architectures.

A central enabling element is the online physical-forward/surrogate-backward training scheme. The forward pass is supplied by the measured optical system, so the optimization directly includes experimental effects such as aberrations, finite resolution, noise, drift and channel nonuniformity. The surrogate is not used as a replacement for the optical processor; it is a differentiable interface that provides gradients for the trainable phase masks. Sharing a single channel-conditioned surrogate across the channel array is important because it preserves the common structure of the optical transformation while allowing channel-specific deviations to be corrected. Online fine-tuning further allows the surrogate to follow the physical system as the optical parameters evolve. This combination of measured forward computation and learned backward differentiation provides a practical route for training large, experimentally realized optical models with more than one million trainable phase parameters.

The present system is intentionally hybrid. Input preparation, camera readout, surrogate learning, learned inter-channel mixing, regression heads and transformer decoding are performed electronically, while the optical stack provides the trainable physical transformation and measured visual representation. This division should not be viewed only as a limitation. It reflects a useful design principle for near-term optical machine learning: optical hardware can perform high-dimensional spatial transformations, while electronic computation supplies calibration, readout, sequence modelling and gradient-based optimization. A system-level contextual comparison of task performance and the trainable phase and electronic parameter counts is provided in Supplementary Section~\ref{sec:supp_results_comparison} and Supplementary Table~\ref{tab:state_of_the_art_comparison}. More broadly, the results of this work suggest that spatial multiplexing should not be viewed solely as a strategy for increasing parallel throughput. Instead, the channel dimension can be incorporated directly into the model design and training process, serving as part of the learned representation itself.

Under the dense full-convolution real-operation-equivalent convention defined for a single optical pass, the present 16-channel processor is estimated to provide \(0.313\) TOPS at an effective \(60\)-Hz optical-pass rate. The corresponding \(30\)-W component-level power estimate gives \(0.0104\) TOPS W\(^{-1}\). The operation-counting convention, power budget and comparison with representative platforms are detailed in Supplementary Section~\ref{sec:supp_energy_efficiency} and Supplementary Tables~\ref{tab:power_consumption} and~\ref{tab:comp_power_consumption}.

Several directions follow naturally from this framework. More expressive and/or fully optical channel-mixing mechanisms could turn the channel axis into a deeper optical feature space. Future implementations could move some of these operations closer to the photonic domain, for example through integrated modulators, optical or optoelectronic channel mixing, faster detector feedback, or learned readout circuits. At the same time, the current free-space platform provides a flexible experimental test bed for exploring how optical degrees of freedom should be organized and trained. Extending the present hybrid vision-language demonstration toward a fully optical vision-language model represents another important direction. Such an architecture would require optical implementations of additional components that are currently electronic, including token generation, sequence processing, attention-like operations, memory mechanisms, and language decoding. While these capabilities remain challenging, advances in integrated photonics, optical interconnects, nonlinear photonic devices, and optoelectronic feedback systems may enable progressively larger portions of the vision-language pipeline to be realized in the optical domain. In the longer term, such developments could move optical processors from trainable visual front ends toward more complete multimodal computing platforms.

\section*{Methods}

\subsection*{Optical setup}

{The optical setup and its characterization are presented extensively in \textit{N\oe{}tinger, Tuuva and Fleury}~\cite{noetinger2026tutorial}; a component-level description is provided in Supplementary Section~\ref{sec:supp_experimental_setup}. It has been explicitly conceived as a prototyping platform to investigate learning algorithms for diffractive neural networks (DNNs) using multiplexing to perform learning and inference in a reasonable amount of time, average noises or provide channel parallelism within the same setup. } \newline
{ A red laser beam at 633nm (Spectra Physics) is expanded using a $4f$-system (lens $L_0$ and $L_0'$). The resulting plane wave is directed to an amplitude modulator (Holoeye microdisplay HEO 2220 CFS with a 4.5\textmu m pixel pitch, referred to as \textit{the microdisplay}) conjugated with a phase-only modulator (Holoeye ERIS with an 8\textmu m pixel pitch, referred to as \textit{the SLM}) using a \textit{4f}-system. The magnification of the $4f$-system is set so that one pixel of the microdisplay corresponds to one pixel of the SLM. The conjugation of the two modulators performs a multiplication of the field from the microdisplay convoluted by point-spread function $h_1$ (taking into account the response of the $4f$-systems from the limited numerical aperture of lenses $L_1$ and $L_2$) and the phase mask on the SLM $\exp\!\left(i2\pi\boldsymbol{\theta}_{\mathrm{phys},c}^{(l)}\right)$. As the two modulators are conjugated, $h_1$ corresponds to an Airy spot, its width is smaller than the size of a SLM pixels \cite{noetinger2026tutorial}. The resulting beam from the SLM is polarization filtered in a conjugate plane and then imaged onto a defocused camera. A diaphragm filters high spatial frequencies in the Fourier plane to reduce the experimental noises. The defocus allows to blend the input data. The amount of \textit{mixing} is set by a tunable parameter $\Delta z$ corresponding to the position of the camera relativley to the conjugate plane, it sets the point-spread function $h_{\Delta z}$. The intensity readout of the camera makes the system non-linear. Thus, all the ingredients of convolutional neural networks are combined.\newline
$N_c=4\times4$ physical regions of interest of $168\times168$ pixels are defined on both the microdisplay and the SLM; within each region, the learned models use an active $112\times112$ computational canvas for the displayed input and trainable phase profile. As it is not possible to align each matching zone manually, each channel would have different transfer function mostly shifted, each zone are aligned thanks to a semi-automatic iterative procedure relying on sharp edge diffraction and automatic ellipse detection. It records the coordinates of the input pixels associated to one channel for the microdisplay and SLM as well as the coordinates of the $N_{out,x}\cdot N_{out,y}$ output pixels on the camera which are saved in a look-up table. $N_b = 8$ acquisitions are performed successively. The interfacing code allows to create an object whose forward method takes the intensity and phasemask as \textit{Pytorch} tensor inputs with shapes $[N_c\cdot N_b,1, N_{in,x}, N_{in,y}]$ or $[N_c, N_b, N_{in,x}, N_{in,y}]$ depending on the applications. The outputs from the camera readout are tensors with shapes $[N_c\cdot N_b,1, N_{out,x}, N_{out,y}]$ or $[ N_b, N_c, N_{out,x}, N_{out,y}]$. The input shape parameters $N_{in,x}$ and $N_{in,y}$ are declared while $N_{out,x}$ and $N_{out,y}$ are measured by the calibration algorithm and recorded. This function is fed to the neural network architecture and surrogate for learning and inferring.}

\section*{Acknowledgments}
\paragraph*{Author contributions:}
A.M. conceived the idea, designed the experiment-based training pipeline, and conducted the theoretical and AI work. G.N. and T.T. carried out the optical experiments. R.F. supervised the project. All authors contributed to the interpretation of the results and the writing of the manuscript.
\paragraph*{Competing interests:}
The authors declare no competing interests.

\paragraph*{Data, code and materials availability:}

The data and code used in this article are available at \href{https://github.com/MomeniAli/Multi-channel-OVM}{MomeniAli/Multi-channel-OVM}.

\clearpage
\normalsize
\renewcommand{\refname}{References}
\bibliography{refs}

\clearpage
\thispagestyle{empty}
\begingroup
\singlespacing
\centering
\vspace*{0.45in}

{\rmfamily\bfseries\fontsize{22}{28}\selectfont Supplementary Material for\par}
\vspace{0.35in}
{\rmfamily\bfseries\fontsize{24}{31}\selectfont Multi-channel Optical Vision Model\par}
\vspace{0.45in}

{\rmfamily\fontsize{14}{20}\selectfont
Ali Momeni\textsuperscript{1}, Guillaume Noetinger\textsuperscript{1}, Tim Tuuva\textsuperscript{1}, and Romain Fleury\textsuperscript{1,*}\par}
\vspace{0.22in}
{\rmfamily\fontsize{11}{16}\selectfont
\textsuperscript{1}Laboratory of Wave Engineering, School of Electrical Engineering, Swiss Federal Institute of Technology in Lausanne (EPFL), Lausanne, Switzerland\par}
\vspace{0.08in}
{\rmfamily\fontsize{11}{16}\selectfont
\textsuperscript{*}Corresponding author. E-mail: romain.fleury@epfl.ch\par}

\vfill
\begin{minipage}{0.86\textwidth}
\raggedright
\hyphenpenalty=10000
\exhyphenpenalty=10000
{\rmfamily\bfseries\fontsize{14}{18}\selectfont This PDF file includes:\par}
\vspace{0.16in}
{\rmfamily\fontsize{12}{18}\selectfont
Supplementary Note 1: Experimental setup\par
Supplementary Note 2: Surrogate model architecture and performance\par
Supplementary Note 3: Task details\par
Supplementary Note 4: Results comparison with state-of-the-art\par
Supplementary Note 5: Energy consumption and TOPS/W\par}
\end{minipage}
\vspace*{0.55in}
\endgroup

\clearpage
\setcounter{section}{0}
\renewcommand{\thesection}{S\arabic{section}}
\renewcommand{\theHsection}{S\arabic{section}}
\renewcommand{\thesubsection}{S\arabic{section}.\arabic{subsection}}
\setcounter{table}{0}
\renewcommand{\thetable}{S\arabic{table}}
\renewcommand{\theHtable}{S\arabic{table}}
\setcounter{figure}{0}
\renewcommand{\thefigure}{S\arabic{figure}}
\renewcommand{\theHfigure}{S\arabic{figure}}
\setcounter{equation}{0}
\renewcommand{\theequation}{S\arabic{equation}}
\renewcommand{\theHequation}{S\arabic{equation}}

\section{Experimental setup}
\label{sec:supp_experimental_setup}

The experimental setup is shown in Supplementary Fig.~\ref{fig:DetailedSetup} and described in detail in Ref.~\cite{noetinger2026tutorial}. \newline

\begin{figure}[b!]
\centering
\includegraphics[width=0.6\linewidth]{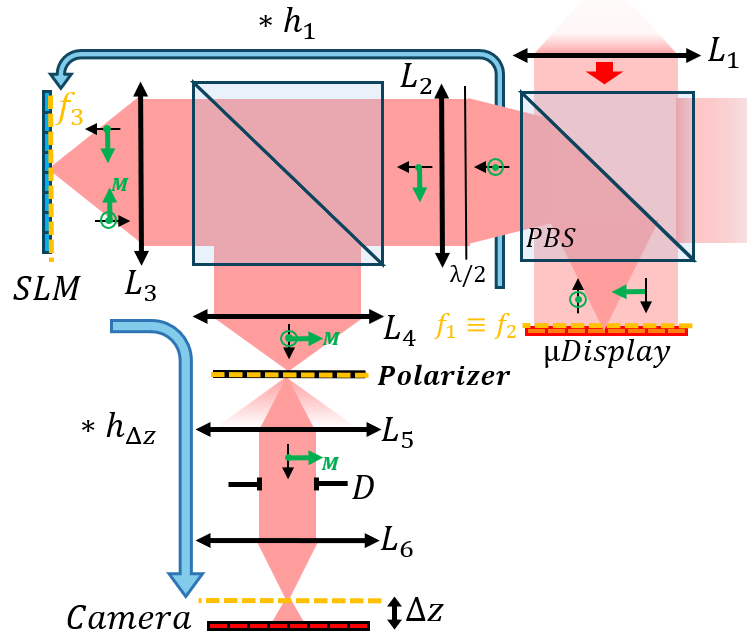}
\caption{\label{fig:DetailedSetup} \textbf{The setup.} An amplitude modulator (the microdisplay, labelled $\mu$Display in the schematic) is conjugated to a phase modulator (the SLM) using a $4f$ system. The resulting field is sent to a defocused camera after filtering by a diaphragm $D$. The polarization is indicated by the green arrows. Only the horizontal polarization is modified by the SLM (green letter $M$). The remaining vertical polarization is filtered with a polarizer. Ideally, the field on the SLM is the convolution of the microdisplay image and the point-spread function $h_1$; propagation to the defocused camera corresponds to convolution by $h_{\Delta z}$. In practice, aberrations, parasitic reflections and cross-talk have to be taken into account.}
\end{figure}

It consists in a horizontally polarized red laser @633nm (Spectra Physics) expanded using a pair of lenses ($L_0$ not shown and $L_1$). The resulting plane wave in the focal plane of $L_1$ is sent to a liquid cristal amplitude modulator (Holoeye microdisplay HEO 2220 CFS with a 4.5\,\textmu m pixel pitch, hereafter referred to as \textit{the microdisplay}). It is conjugated with a phase-only modulator (Holoeye ERIS with an 8\,\textmu m pixel pitch, hereafter referred to as \textit{the SLM}) using a \textit{4f}-system made of lenses $L_2$ and $L_3$. The magnification of this \textit{4f}-system is adjusted so that roughly one pixel of the microdisplay corresponds to one pixel of the SLM. This is achieved by choosing two lenses with the right focal length ratio and by slightly defocusing the $4f$-system, this defocus mainly adds field curvature in the plane of the SLM but it is negligible with respect to the size of the images displayed (roughly $100\cdot100$ px$^2$). The resulting beam is therefore modulated both in phase and in amplitude although it includes aberrations from the non-perfect modulator surface \cite{Moser:19}, the defocus and all the other optics. Only one polarization is modulated by the SLM hence we have to make sure that light arrives with the right polarization on the SLM. The microdisplay only modulates horizontally polarized incoming light and then reflects a vertically polarized light which can be collected using a polarizing beam splitter (PBS) cube on another port. Then, we use a $\lambda/2$ plate at $45°$ after the microdisplay to make sure the polarization becomes horizontal before going to the SLM. The modulated beam is directed in another direction thanks to a beamsplitter. The remaining horizontally polarized light is filtered in a conjugated plane made with the $4f$-system of lenses $L_3$ and $L_4$. The beam is then sent to a defocused camera by the $4f$-system $L_5$-$L_6$. Along this path, high spatial frequencies are filtered using a diaphragm in the Fourier plane of the microdisplay and SLM. We define the point-spread functions from the microdisplay to the SLM as $h_1$ and from the SLM to the camera as $h_{\Delta z}$.
A grid of $4\times 4$ independent regions of interest (RoIs) of size $N_{in,x}\cdot N_{in,y}=168 \times 168$ pixels each is used on the microdisplay and the SLM. They are separated by more than 100 pixels to ensure there is no crosstalk between them. This corresponds to more than $4\cdot 10^5$ degrees of freedom. To reach a pixel wise alignment between each RoI of the microdisplay and SLM so that each channel performs approximately the same physical operation we next use an automatic procedure. This procedure on displaying ellipses in amplitudes on the microdisplay and detecting their position thanks to a standard ellipse detection algorithm from the Python \textit{cv2} package, displaying ellipses of phase contrast \cite{noetinger2026tutorial} and detecting their positions, shifting the first ones relatively to the other iteratively so that the detected center finally match. This is robust to defocus and large angular or translation mismatch. This algorithm also provides the size ($N_{out,x}\cdot N_{out,y}$) and position of the corresponding output zone on the camera. The corresponding code is available on GitHub [\url{https://github.com/TTimTT/SLM-alignment}]. \newline
An interfacing code allows to build a function requesting two $[N_c, N_b, N_{in,x}, N_{in,y}]$ as inputs with $N_c$ the number of spatial channels (=RoIs) and $N_b$ a number corresponding to a succession of shots taken consecutively to speed the process. Its output is then a tensor $[N_c, N_b, N_{out,x}, N_{out,y}]$. This base function is used in th \textit{forward} method of each layer.

\section{Surrogate model architecture and performance}
\label{sec:supp_surrogate_model}

The surrogate model provides the differentiable counterpart of the measured optical forward pass used during physical-forward/surrogate-backward training. Its role is not to replace the optical processor during the forward computation, but to provide gradients with respect to the displayed phase profiles while preserving the experimentally measured output as the value from which the task loss is computed. We therefore designed the surrogate as a hybrid physics-informed and data-driven model. The physics-informed part is the single-channel optical transformation in Eq.~\ref{eq:physical_channel_forward}: the displayed input field is modulated by a phase mask, propagated through effective point-spread functions, converted to intensity at the camera plane and then cropped and normalized. In the code, this single-channel propagation prior is represented by the PSF-based module \texttt{PSFNet}, which applies complex phase modulation, FFT-based convolution with a cached PSF, intensity detection and output normalization. This component defines the optical structure that the surrogate is asked to approximate.

The full surrogate used in the experiments, \texttt{Surrogate\_OpticalNet\_Unet}, augments this physical prior with learned corrections that account for effects not captured by a single idealized propagation kernel. These include channel-dependent aberrations, small alignment differences, parasitic reflections, speckle-like spatial structure, camera readout variations and residual mismatch between the calibrated optical model and the actual hardware. The architecture shown in Fig.~\ref{fig:fig2}a follows this logic. The input intensity pattern and the trainable phase mask are first mapped to a common optical canvas. The phase is represented by multiple sinusoidal harmonics, while explicit spatial coordinates are added so that the network can learn position-dependent optical and detection effects. A shared encoder-decoder then captures the dominant propagation response, and channel-conditioned residual modules refine the prediction for each of the 16 spatial channels. Table~\ref{tab:surrogate_architecture} gives the full structural and architectural decomposition of the surrogate, including the physics-informed prior, shared encoder-decoder and channel-conditioned correction paths.

\begin{center}
\scriptsize
\setlength{\tabcolsep}{3pt}
\renewcommand{\arraystretch}{1.23}
\begin{longtable}{>{\raggedright\arraybackslash}p{0.15\textwidth} >{\raggedright\arraybackslash}p{0.22\textwidth} >{\raggedright\arraybackslash}p{0.27\textwidth} >{\raggedright\arraybackslash}p{0.25\textwidth}}
\caption{\textbf{Technical decomposition of the surrogate architecture in Fig.~\ref{fig:fig2}a.} The table maps the code-level implementation to the optical function of each module. The surrogate combines a single-channel physics-informed propagation prior with a shared encoder-decoder and several channel-conditioned residual paths that account for the measured differences between the 16 optical tiles.}
\label{tab:surrogate_architecture}\\
\toprule
\textbf{Subsystem} & \textbf{Code modules} & \textbf{Implemented operation} & \textbf{Role in the surrogate} \\
\midrule
\endfirsthead
\toprule
\textbf{Subsystem} & \textbf{Code modules} & \textbf{Implemented operation} & \textbf{Role in the surrogate} \\
\midrule
\endhead
Physics-informed optical prior & \texttt{PSFNet} & Applies complex phase modulation, FFT-based convolution with a cached PSF, camera-plane intensity detection, center cropping and per-sample normalization. & Implements the Eq.~\ref{eq:physical_channel_forward}-type single-channel propagation used as the physical reference for the learned surrogate. \\
Input and phase alignment & \texttt{encoding\_x\_phase\_}\newline\texttt{physical}; \texttt{x\_upsampler}; \texttt{phase\_upsampler}; \texttt{LearnedUpsampler} & Each branch uses shallow convolutional refinement, bilinear resizing and post-resize refinement to place the input image and phase mask on the common \(112\times112\) encoder canvas. & Aligns the two experimentally controlled quantities before joint propagation modeling and reduces artifacts that would arise from a purely geometric resize. \\
Periodic phase representation & \texttt{phase\_harmonics=20} & Converts the scalar phase into 40 harmonic channels, \(\{\cos(m\phi),\sin(m\phi)\}_{m=1}^{20}\), with \(\phi=2\pi\) phase. Together with the input and coordinates, this gives a 44-channel encoder input. & Represents the periodic dependence of the optical field on the displayed phase, which is not naturally captured by a single scalar phase channel. \\
Spatial coordinate conditioning & CoordConv channels \((x,y,r)\) & Appends horizontal, vertical and radial coordinate maps on the encoder canvas. & Allows the surrogate to model field-position effects, including nonuniform illumination, channel-dependent aberrations and camera-plane distortions. \\
Residual encoder & \texttt{enc1}, \texttt{enc2}, \texttt{enc3}; \texttt{ResidualBlock3x3}; \texttt{SafeGroupNorm}; GELU & Three stride-2 residual stages map \(112\times112\rightarrow56\times56\rightarrow28\times28\rightarrow14\times14\), with feature widths \(45\rightarrow90\rightarrow180\). Each block contains two 3x3 convolutions and a 1x1 residual skip when the scale or channel width changes. & Extracts the shared optical response while progressively increasing the receptive field, so that local input-phase interactions can be integrated into larger camera-plane structures. \\
Shared bottleneck and attention & \texttt{bottleneck}; \texttt{BottleneckDilated}; \texttt{SEBlock}; \texttt{SpatialAttention} & Operates at \(14\times14\) with 180 channels. The bottleneck combines a residual block, dilated 3x3 convolutions with dilation factors 1, 2 and 3, channel-wise squeeze-excitation and spatial attention. & Models the long-range redistribution of optical intensity and adaptively reweights feature channels and spatial regions that are important for reconstructing the measured output. \\
Coarse channel residual & \texttt{channel\_bottleneck\_}\newline\texttt{blocks}; \texttt{ChannelBottleneck}\newline\texttt{Block}; \texttt{chan\_ids} & Applies one channel-specific 1x1 residual block at the bottleneck scale; the correct block is selected by the channel index. The residual is initialized with a small learnable scale. & Introduces low-resolution, channel-dependent corrections before decoding, capturing coarse aberration and alignment differences between optical tiles. \\
Decoder and skip fusion & \texttt{dec3\_up}, \texttt{dec2\_up}, \texttt{dec1\_up}; \texttt{PixelShuffle}; \texttt{fuse3}, \texttt{fuse2}, \texttt{fuse1}; \texttt{dec3\_rb}, \texttt{dec2\_rb}, \texttt{dec1\_rb} & PixelShuffle stages reconstruct \(14\times14\rightarrow28\times28\rightarrow56\times56\rightarrow112\times112\). Encoder features are concatenated at matching scales and fused with 1x1 convolutions before residual refinement. & Recovers camera-plane spatial detail while preserving the high-resolution input, phase-harmonic and coordinate information through U-Net-like skip connections. \\
Mid-resolution channel residual & \texttt{channel\_mid\_blocks}; \texttt{ChannelMidResBlock} & Applies one channel-specific nonlinear 3x3 residual CNN at the \(56\times56\) decoder scale, with a small learnable residual gain. & Refines intermediate spatial structure for each tile, where channel-dependent texture, blur and speckle-like deviations are more visible than at the bottleneck scale. \\
Shared intensity head & \texttt{final}; depthwise refinement; \texttt{SoftClip01} & Maps the final decoder features to one intensity channel using 3x3, 1x1, depthwise 3x3 and 1x1 convolutions, followed by a smooth clipping function with \(\beta=3\). & Produces the shared camera-intensity estimate and constrains the prediction to a physically meaningful range before channel-conditioned output correction. \\
Channel embedding and spatial basis & \texttt{channel\_emb}; \texttt{spatial\_basis}; layer normalization; \texttt{einsum} & Uses a 512-dimensional embedding for each of the 16 channels and combines it with a learned spatial basis of size \(512\times112\times112\) to form a one-channel spatial conditioning map. & Encodes the identity of each optical tile as a structured spatial field, rather than as a scalar label, allowing static tile-specific deviations to condition the final prediction. \\
Output-level channel correction & \texttt{ChannelHead} & Concatenates the shared intensity estimate with the channel-conditioning map and predicts an additive residual using local 3x3 convolution, dilated 3x3 convolution, depthwise 3x3 convolution and 1x1 bottleneck fusion. & Corrects the final camera-plane prediction for channel-specific residual errors while retaining a shared correction rule across channels. \\
Target matching and model scale & \texttt{\_match\_to\_target}; \texttt{param\_stats} & Resizes or center pad-crops the corrected output to the measured target size. & Aligns the predicted image with the measured camera output and reflects the design choice that a shared propagation backbone should be complemented by substantial channel-conditioning capacity. \\
\bottomrule
\end{longtable}
\end{center}

For the experiments reported in the main text, the surrogate is shared across all 16 optical channels. The saved model uses a \(112\times112\) encoder canvas, a base width of 45 feature channels, 20 phase harmonics and a 512-dimensional channel embedding. The encoder input therefore contains the optical input intensity, 40 phase-harmonic channels and three coordinate channels. This design makes the surrogate neither a set of 16 independent networks nor a purely channel-agnostic model. Instead, most of the propagation structure is learned in a shared pathway, while channel-conditioned residual blocks supply the corrections required by individual optical tiles. These corrections enter at the bottleneck, at the mid-resolution decoder scale and at the output plane, where a learned channel embedding is projected through a spatial basis and used to condition the final residual prediction.

The surrogate is first pre-trained against measured optical input-output pairs. During this stage, the optical system is queried with randomized input-field and phase-mask pairs, and the measured camera output is used as the target. The random fields are generated on the optical canvas by filtering uniform noise with a randomly selected spatial kernel, including identity, isotropic or anisotropic Gaussian, signed random, difference-of-Gaussians and Laplacian kernels, followed by per-sample normalization. This procedure exposes the surrogate to both slowly varying intensity distributions and edge-rich spatial structures, thereby sampling a broad range of spatial frequencies before task-specific optimization. For a predicted optical response \(\hat{\mathbf{y}}\) and measured response \(\mathbf{y}\), the surrogate was optimized with the weighted objective
\begin{equation}
\mathcal{L}_{\mathrm{sur}} =
\alpha_1\mathcal{L}_{\mathrm{MSE}}(\hat{\mathbf{y}},\mathbf{y})
+\alpha_2\mathcal{L}_{\mathrm{MS\text{-}SSIM}}(\hat{\mathbf{y}},\mathbf{y})
+\alpha_3\mathcal{L}_{\mathrm{FFL}}(\hat{\mathbf{y}},\mathbf{y}).
\label{eq:surrogate_loss}
\end{equation}
Here, \(\mathcal{L}_{\mathrm{MSE}}\) is the pixelwise mean-squared error and constrains the absolute camera-plane intensity. The structural term is defined as \(\mathcal{L}_{\mathrm{MS\text{-}SSIM}}=1-\mathrm{MS\text{-}SSIM}\), and penalizes differences in luminance, contrast and spatial structure across multiple image scales. The MS-SSIM term used the five-scale weights \((0.0448,0.2856,0.3001,0.2363,0.1333)\), renormalized only when fewer scales were required by the output size. Finally, \(\mathcal{L}_{\mathrm{FFL}}\) denotes the focal-frequency loss, which compares the prediction and measurement in the Fourier domain and emphasizes spectral components that are poorly reconstructed. This combination penalizes pixelwise intensity error while also constraining structural similarity and spatial-frequency content. Representative examples shown in Fig.~\ref{fig:fig2}a reach a generated-image MSE of \(5.1\times10^{-4}\).

\begin{figure}[p]
\centering
\includegraphics[height=0.9\textheight]{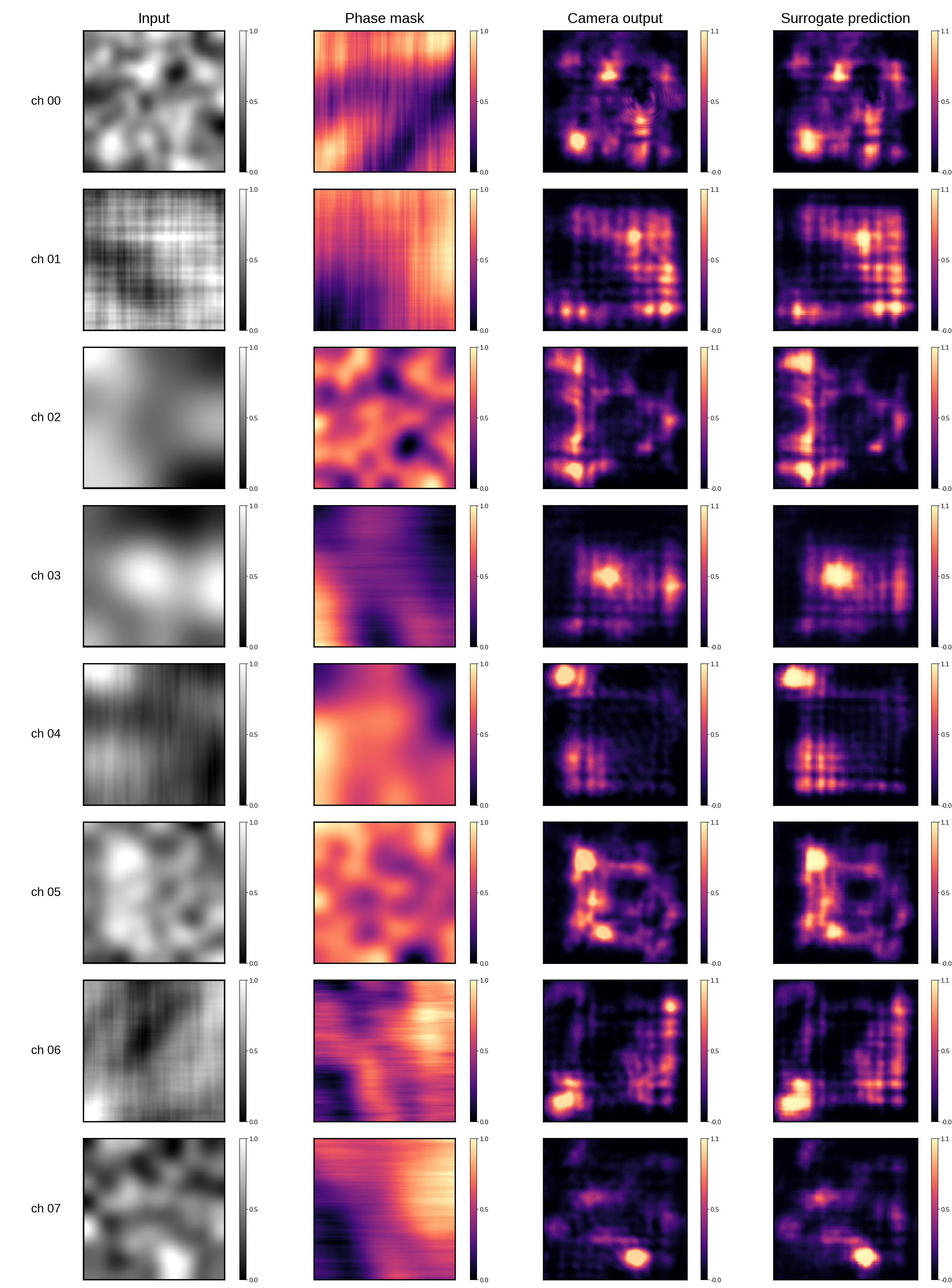}
\caption{\textbf{Surrogate prediction examples for channels 0-7.} Randomized input fields and phase masks are shown together with the corresponding measured camera outputs and surrogate predictions.}
\label{fig:surrogate_channel_examples_a}
\end{figure}

\begin{figure}[p]
\centering
\includegraphics[height=.9\textheight]{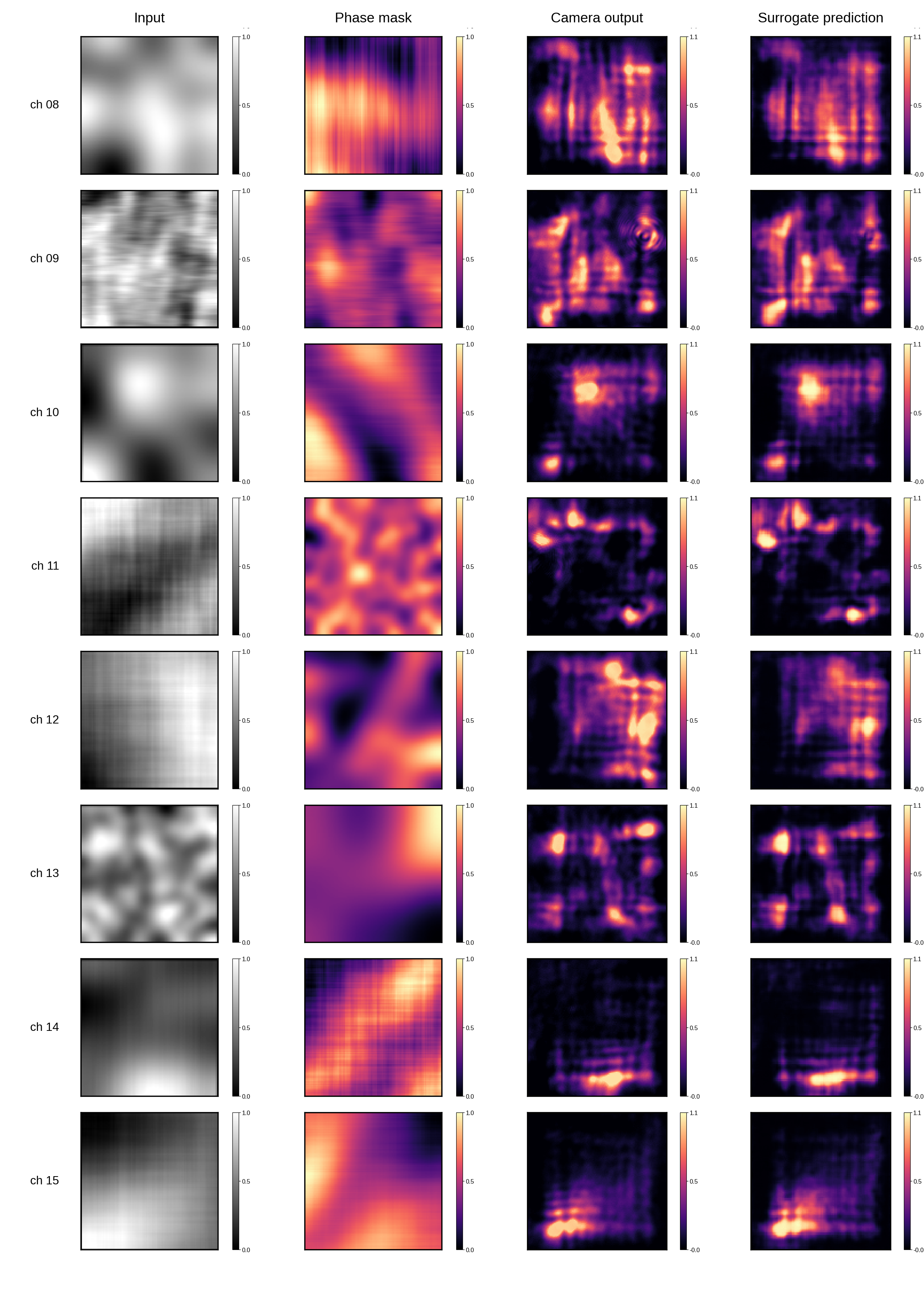}
\caption{\textbf{Surrogate prediction examples for channels 8-15.} The remaining optical channels show the same comparison between randomized input fields, phase masks, measured camera outputs and surrogate predictions.}
\label{fig:surrogate_channel_examples_b}
\end{figure}

Figures~\ref{fig:surrogate_channel_examples_a} and \ref{fig:surrogate_channel_examples_b} provide a channel-resolved view of this pre-training performance across the full 16-channel optical field of view. Each row corresponds to one spatial channel, and the four columns show, from left to right, the randomized input field, the applied phase mask, the experimentally measured camera output and the surrogate prediction. The 16 channels are split into two panels, with channels 0-7 in Fig.~\ref{fig:surrogate_channel_examples_a} and channels 8-15 in Fig.~\ref{fig:surrogate_channel_examples_b}, to make the individual channel comparisons readable. Together, Figs.~\ref{fig:surrogate_channel_examples_a} and \ref{fig:surrogate_channel_examples_b} show that the surrogate model accurately predicts the measured output of the optical setup across the spatially multiplexed channel set.

\clearpage

\section{Task details}
\label{sec:supp_task_details}

Table~\ref{tab:task_interface} summarizes the task interface used for each experiment, including the dataset scale, optical encoding and output definition. Table~\ref{tab:optimization_protocol} then summarizes the corresponding optimization objectives, optimized variables and validation metrics.

\begin{center}
\scriptsize
\setlength{\tabcolsep}{4pt}
\renewcommand{\arraystretch}{1.25}
\begin{longtable}{>{\raggedright\arraybackslash}p{0.20\textwidth} >{\raggedright\arraybackslash}p{0.34\textwidth} >{\raggedright\arraybackslash}p{0.37\textwidth}}
\caption{\textbf{Task interface and optical readout definitions.} Each task is specified by the dataset and optical encoding used at the input side, and by the mathematical definition of the target or readout used at the output side.}
\label{tab:task_interface}\\
\toprule
\textbf{Task} & \textbf{Dataset and optical encoding} & \textbf{Output target / readout definition} \\
\midrule
\endfirsthead
\toprule
\textbf{Task} & \textbf{Dataset and optical encoding} & \textbf{Output target / readout definition} \\
\midrule
\endhead
Independent-channel MNIST classification &
MNIST grayscale digits, originally \(28\times28\), expanded onto the \(112\times112\) optical canvas using deterministic four-rotation input tiling. The 16 optical channels are trained as parallel channel instances without inter-channel digital mixing. &
The target is the digit label \(y\in\{0,\ldots,9\}\). For the output intensity \(Y_b\) of channel \(b\), the tile score for class \(k\) is
\(s_{b,k}=|M_k|^{-1}\sum_{h,w}M_k(h,w)Y_b(h,w)/\sum_{h,w}Y_b(h,w)\),
where \(M_k\) is the output-plane mask for class \(k\). \\
Class-code MNIST readout &
MNIST grayscale digits, originally \(28\times28\), encoded on the \(112\times112\) optical canvas using deterministic four-rotation input tiling. The 16 optical channels act as binary code channels. &
The target is the digit label \(y\in\{0,\ldots,9\}\), represented by a fixed class-code matrix \(C\in\{-1,+1\}^{10\times16}\). For code channel \(b\), YES and NO mask energies \(s_b^{+}\) and \(s_b^{-}\) define the margin
\(m_b=\log(s_b^{+}+\epsilon)-\log(s_b^{-}+\epsilon)\).
The class logit is \(z_k=16^{-1/2}\sum_{b=1}^{16}C_{k,b}m_b\). \\
Fashion-MNIST channel mixing &
Fashion-MNIST grayscale images, originally \(28\times28\), mapped to 16 channels on the \(112\times112\) optical canvas. Between optical layers, learned digital \(16\rightarrow16\) channel mixers combine information across channels. &
The target is the clothing label \(y\in\{0,\ldots,9\}\). Each channel produces class scores \(s_{b,k}\) from the same normalized tile-energy readout. The channel logits \(z_{b,k}\) are log-scaled and fused as
\(z_k=\sum_b \mathrm{softmax}(\mathbf{a})_b z_{b,k}\),
where \(\mathbf{a}\) are learned channel-fusion weights. \\
Facial-keypoint regression &
Facial keypoint images, \(96\times96\) grayscale, patchified into a \(4\times4\) grid of 16 overlapping optical channels. Each channel is processed on the optical phase/camera canvas. &
The target is a normalized coordinate vector \(\mathbf{p}\in[0,1]^{30}\), corresponding to 15 facial keypoints. From the optical output \(Y_b\), a central mask \(M\) is applied and average-pooled to form \(\mathbf{f}_b\in\mathbb{R}^{169}\). Each channel predicts
\(\hat{\mathbf{p}}_b=\mathrm{ReLU}(W_b\mathbf{f}_b+\mathbf{c}_b)\),
and the final prediction is \(\hat{\mathbf{p}}=\sum_b \mathrm{softmax}(\mathbf{a})_b\hat{\mathbf{p}}_b\). \\
vision-language captioning &
Synthetic composite images generated from MNIST, Fashion-MNIST and NotMNIST families on a \(112\times112\) RGB canvas. The image is divided into a \(4\times4\) array of 16 overlapping optical channels and processed by a six-layer optical encoder with inter-layer channel mixing. &
The target is a tokenized caption sequence \(\mathbf{w}=(w_1,\ldots,w_T)\). The final optical output is converted by a convolutional patch-token readout into memory tokens \(\mathbf{M}\in\mathbb{R}^{64\times256}\), which condition a transformer decoder through cross-attention:
\(p(\mathbf{w}|X)=\prod_t p(w_t\,|\,w_{<t},\mathbf{M})\). \\
\bottomrule
\end{longtable}
\end{center}

\begin{center}
\scriptsize
\setlength{\tabcolsep}{3pt}
\renewcommand{\arraystretch}{1.25}
\begin{longtable}{>{\raggedright\arraybackslash}p{0.17\textwidth} >{\raggedright\arraybackslash}p{0.37\textwidth} >{\raggedright\arraybackslash}p{0.22\textwidth} >{\raggedright\arraybackslash}p{0.18\textwidth}}
\caption{\textbf{Optimization and validation protocol.} The table lists the task-specific training objective, the optimized model variables and the metrics computed on the held-out validation subset for each trained optical model. Here, \(\mathrm{CE}\) denotes cross-entropy and \(\mathrm{BCE}\) denotes binary cross-entropy.}
\label{tab:optimization_protocol}\\
\toprule
\textbf{Task} & \textbf{Training objective} & \textbf{Optimized variables} & \textbf{Optimizer and validation} \\
\midrule
\endfirsthead
\toprule
\textbf{Task} & \textbf{Training objective} & \textbf{Optimized variables} & \textbf{Optimizer and validation} \\
\midrule
\endhead
Independent-channel MNIST classification &
For channel \(b\), logits are obtained from the tile scores as \(z_{b,k}=\tau_b\log(s_{b,k}+\epsilon)\). The objective combines channel-wise classification, energy separation and learnable channel uncertainty:
\(\mathcal{L}=\sum_b[\mathrm{CE}(\mathbf{z}_b,y)/(2\sigma_b^2)+\log\sigma_b]+\lambda_m\mathcal{L}_{\mathrm{margin}}\),
with \(\lambda_m=0.15\). &
Five optical phase-mask layers, per-channel readout temperature \(\tau_b\) and channel uncertainty \(\sigma_b\). &
AdamW with cosine decay, learning rate \(7\times10^{-4}\) to \(1\times10^{-5}\). Validation metrics are mean loss, mean accuracy and per-channel accuracy. \\
Class-code MNIST readout &
The class-code logits \(z_k\) are trained with decoded class loss, auxiliary code-bit supervision, decoded-margin regularization and spillover suppression:
\(\mathcal{L}=\mathrm{CE}(\mathbf{z},y)+0.05\,\mathrm{BCE}(\mathbf{m},\mathbf{c}_y)+0.05\,\mathcal{L}_{\mathrm{margin}}+0.05\,\mathcal{L}_{\mathrm{spill}}\),
where \(\mathbf{c}_y\) is the target code for class \(y\). &
Five optical phase-mask layers and per-channel readout temperature. &
AdamW with cosine decay, learning rate \(2\times10^{-4}\) to \(1\times10^{-6}\). Validation metrics are decoded classification loss and 10-class accuracy. \\
Fashion-MNIST channel mixing &
The fused class logit is \(z_k=\sum_b\mathrm{softmax}(\mathbf{a})_b z_{b,k}\). The training objective is
\(\mathcal{L}=\mathrm{CE}(\mathbf{z},y)+0.05\,\mathcal{L}_{\mathrm{margin}}\),
where the margin term encourages separation between the target and strongest competing class. &
Five optical phase-mask layers, input adapter, inter-layer \(16\rightarrow16\) mixers, channel-fusion weights and readout temperature. &
AdamW with cosine decay, learning rate \(1.06\times10^{-4}\) to \(1\times10^{-6}\). Validation metrics are fused loss, fused accuracy and per-channel accuracy. \\
Facial-keypoint regression &
For normalized keypoints \(\mathbf{p}\), the objective combines coordinate regression with suppression of optical power outside the central readout region:
\(\mathcal{L}=\sqrt{\mathrm{MSE}(\hat{\mathbf{p}},\mathbf{p})+10^{-12}}+0.8\,\|Y\odot(1-M)\|_2^2\). &
Five optical phase-mask layers, inter-layer channel mixers, per-channel linear readout heads and learnable channel-averaging weights. &
AdamW with cosine decay, learning rate \(2\times10^{-4}\) to \(1\times10^{-5}\). Validation metrics are loss and coordinate MSE. \\
vision-language captioning &
Caption tokens are trained by teacher forcing with label-smoothed cross-entropy over non-padding positions \(\Omega\):
\(\mathcal{L}=-|\Omega|^{-1}\sum_{t\in\Omega}\sum_v q_{0.1}(v|w_t)\log p_\theta(v|w_{<t},\mathbf{M})\). &
Six optical phase-mask layers, inter-layer \(16\rightarrow16\) mixers, patchify input projector, convolutional patch-token readout and transformer decoder. &
AdamW with cosine decay, learning rate \(1\times10^{-4}\) to \(1\times10^{-5}\). Validation metrics are caption cross-entropy and BLEU-1 through BLEU-4 from beam-search generation. \\
\bottomrule
\end{longtable}
\end{center}

\begin{figure}[p]
\centering
\includegraphics[height=0.78\textheight]{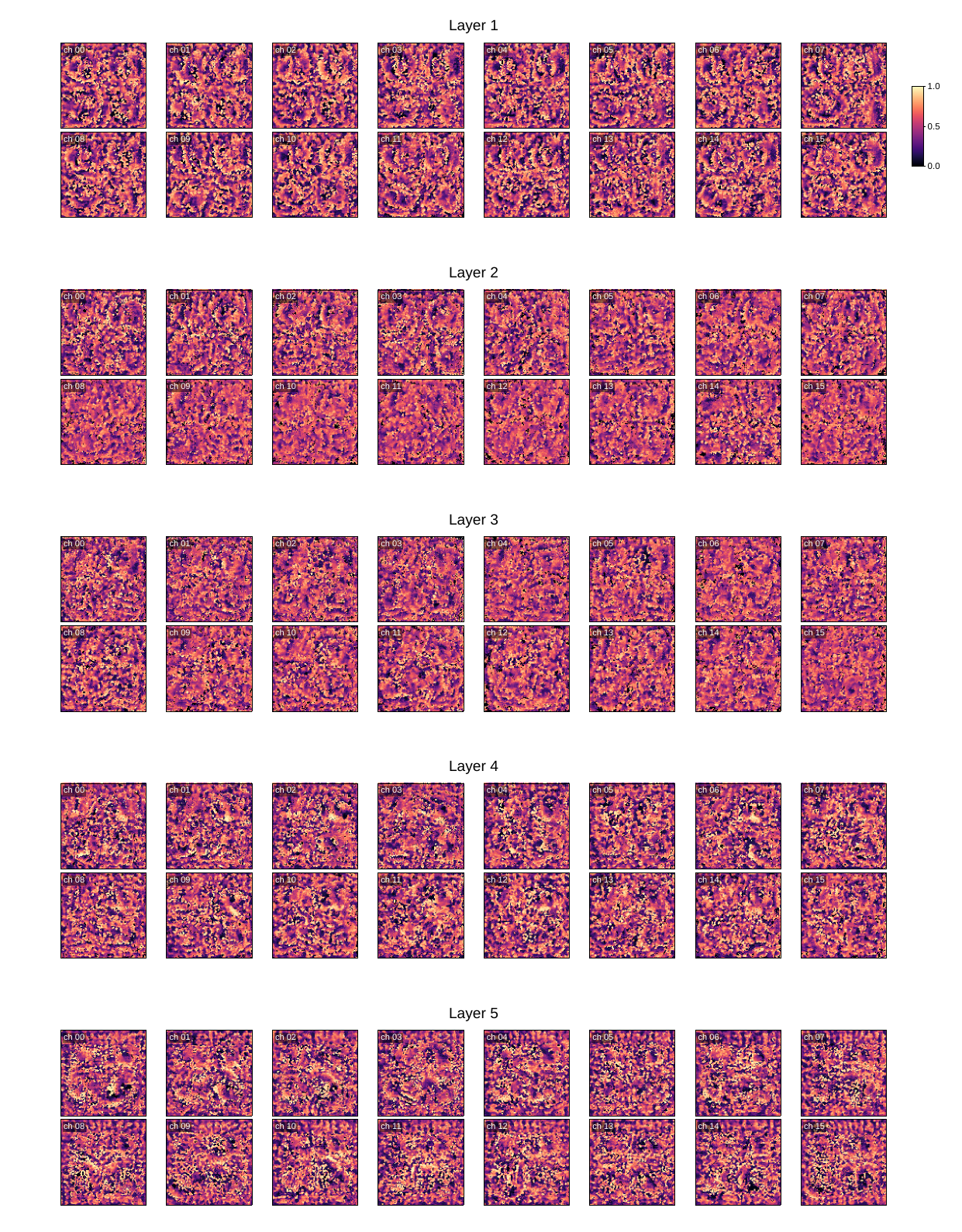}
\caption{\textbf{Trained phase masks for independent-channel MNIST classification.} The optimized phase masks are shown for the five optical layers and all 16 spatial channels (ch~00--ch~15). Each channel contains an independently trainable phase profile and acts as a parallel MNIST classifier without inter-channel mixing.}
\label{fig:phase_masks_independent_mnist}
\end{figure}

\begin{figure}[p]
\centering
\includegraphics[height=0.78\textheight]{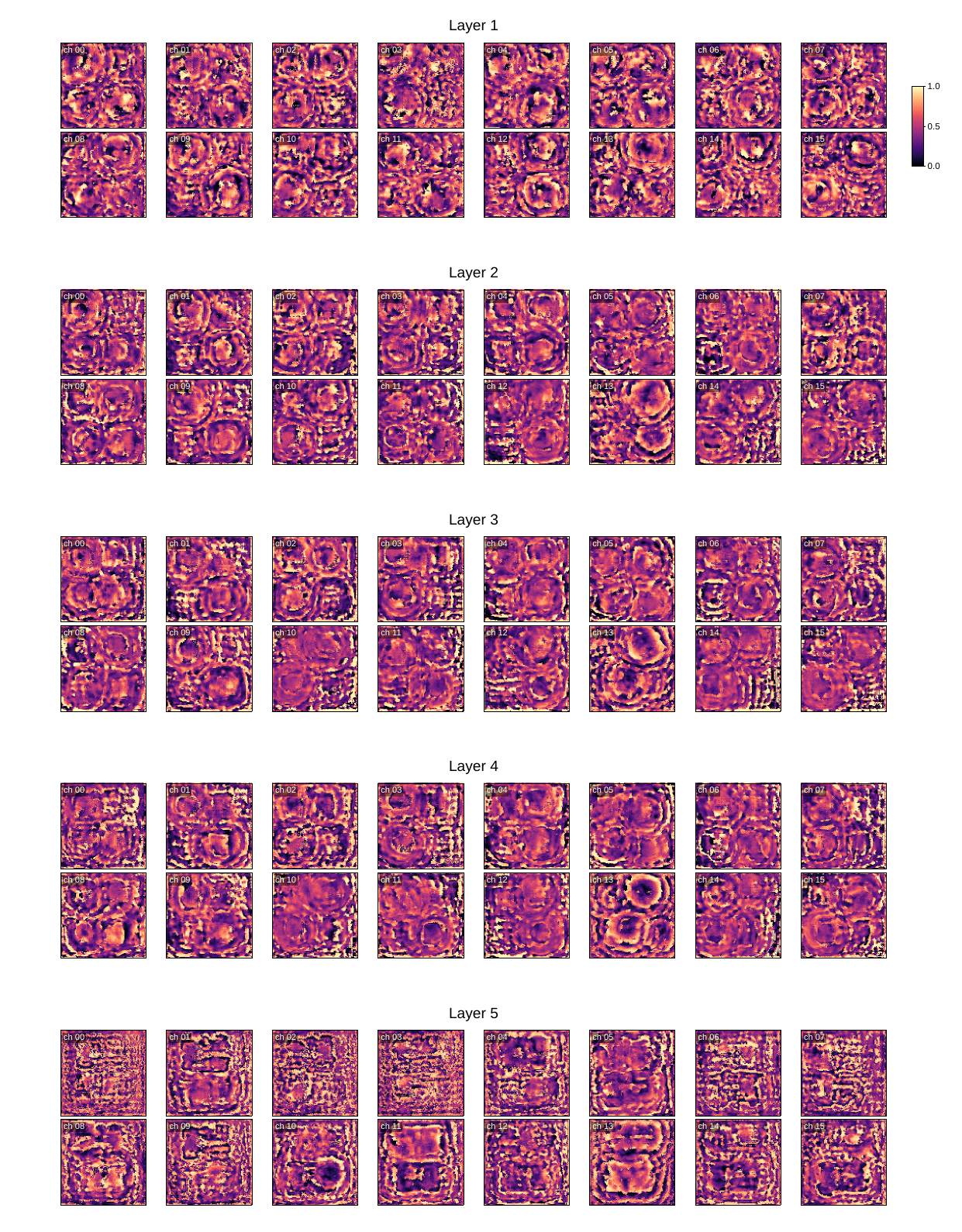}
\caption{\textbf{Trained phase masks for class-code MNIST readout.} The optimized phase masks are shown for the five optical layers and 16 code channels. Each channel generates a binary optical response.}
\label{fig:phase_masks_class_code}
\end{figure}

\begin{figure}[p]
\centering
\includegraphics[height=0.78\textheight]{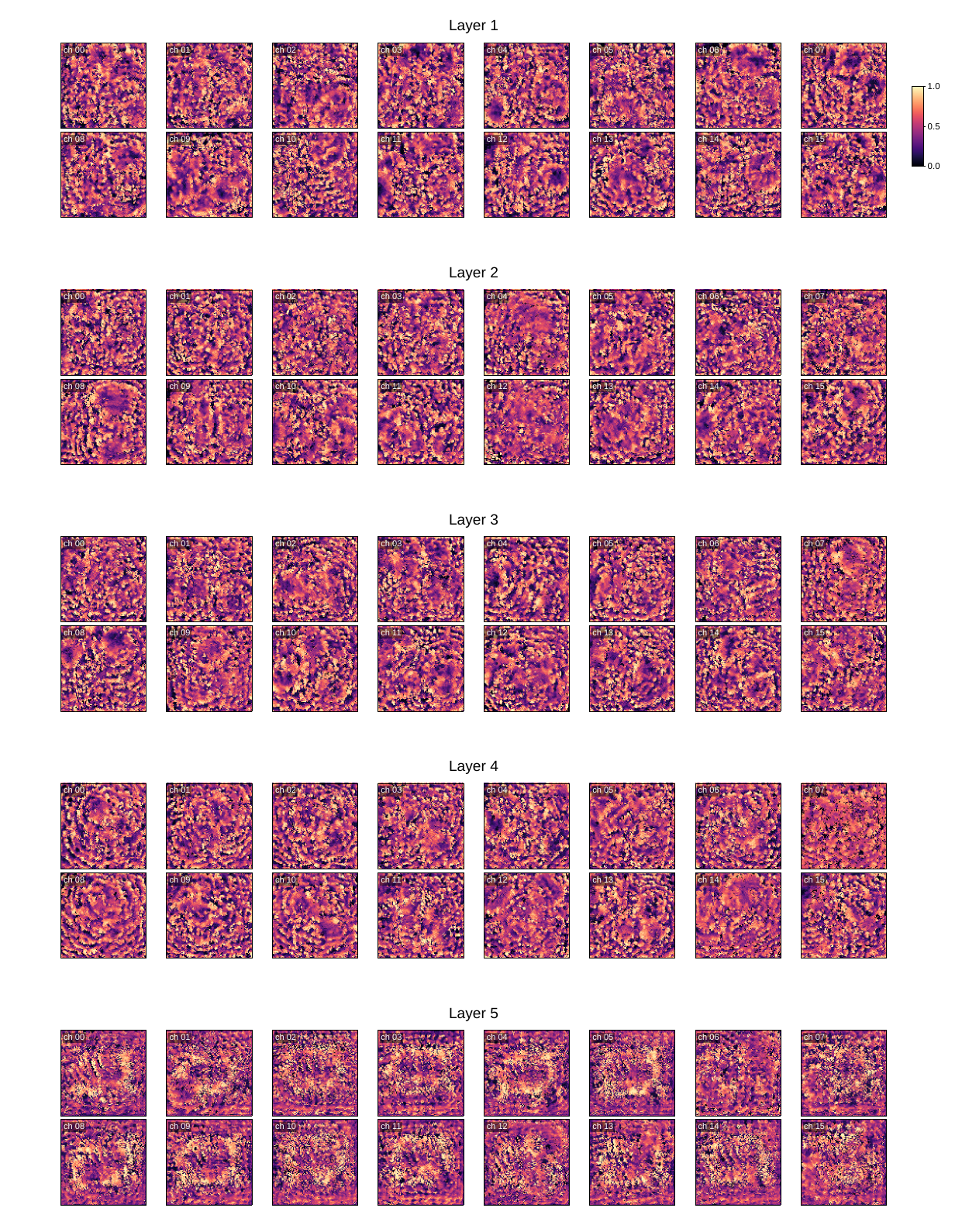}
\caption{\textbf{Trained phase masks for Fashion-MNIST channel mixing.} The optimized phase masks are shown for the five optical layers and all 16 spatial channels. The phase masks are trained jointly with the learned inter-layer \(16\rightarrow16\) channel mixers and the channel-fusion readout, enabling information from different channel features to interact across successive optical layers.}
\label{fig:phase_masks_fashion_mixing}
\end{figure}

\begin{figure}[p]
\centering
\includegraphics[height=0.78\textheight]{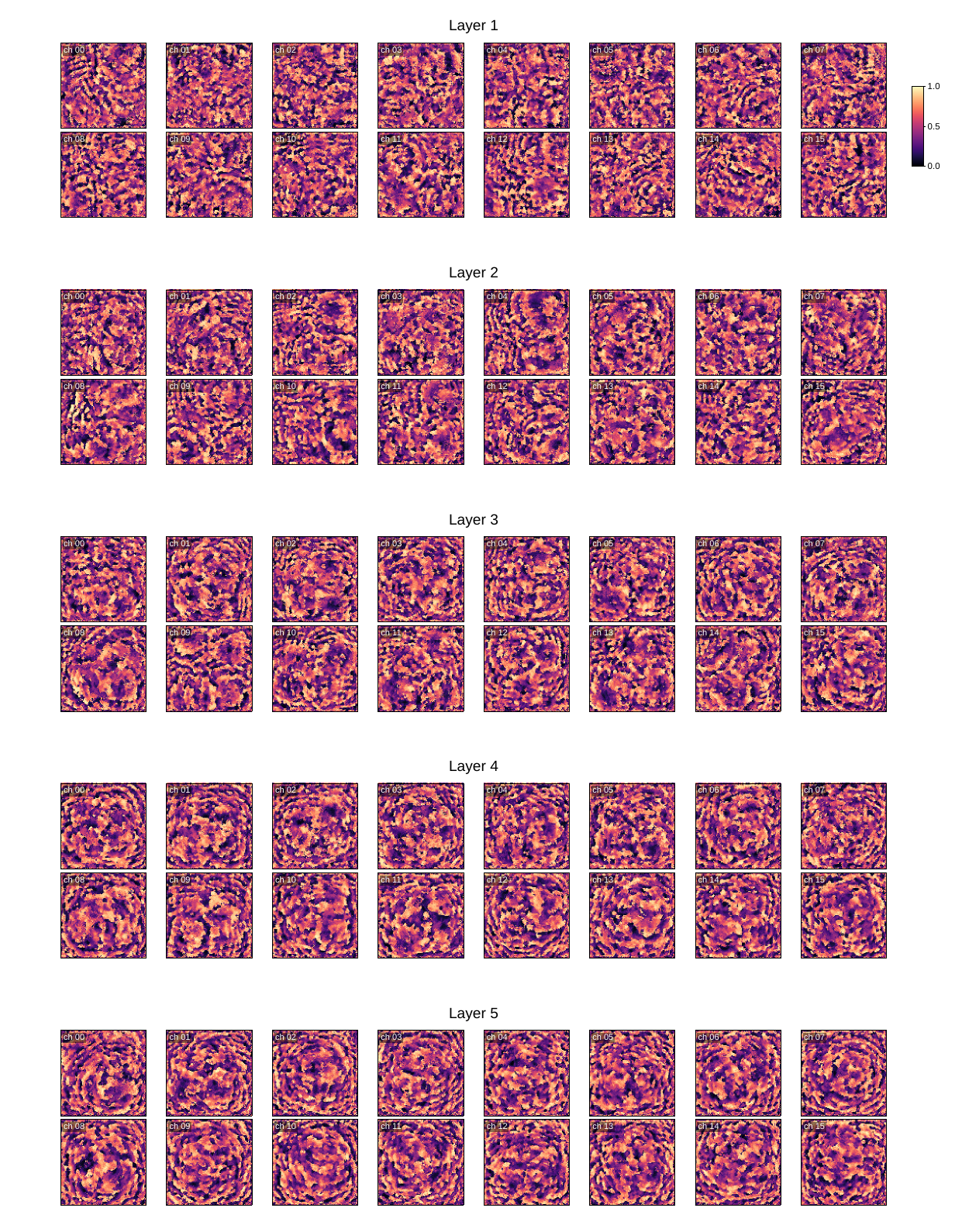}
\caption{\textbf{Trained phase masks for facial-keypoint regression.} The optimized phase masks are shown for the five optical layers and all 16 spatial channels.}
\label{fig:phase_masks_face_regression}
\end{figure}

\begin{figure}[p]
\centering
\includegraphics[height=0.78\textheight]{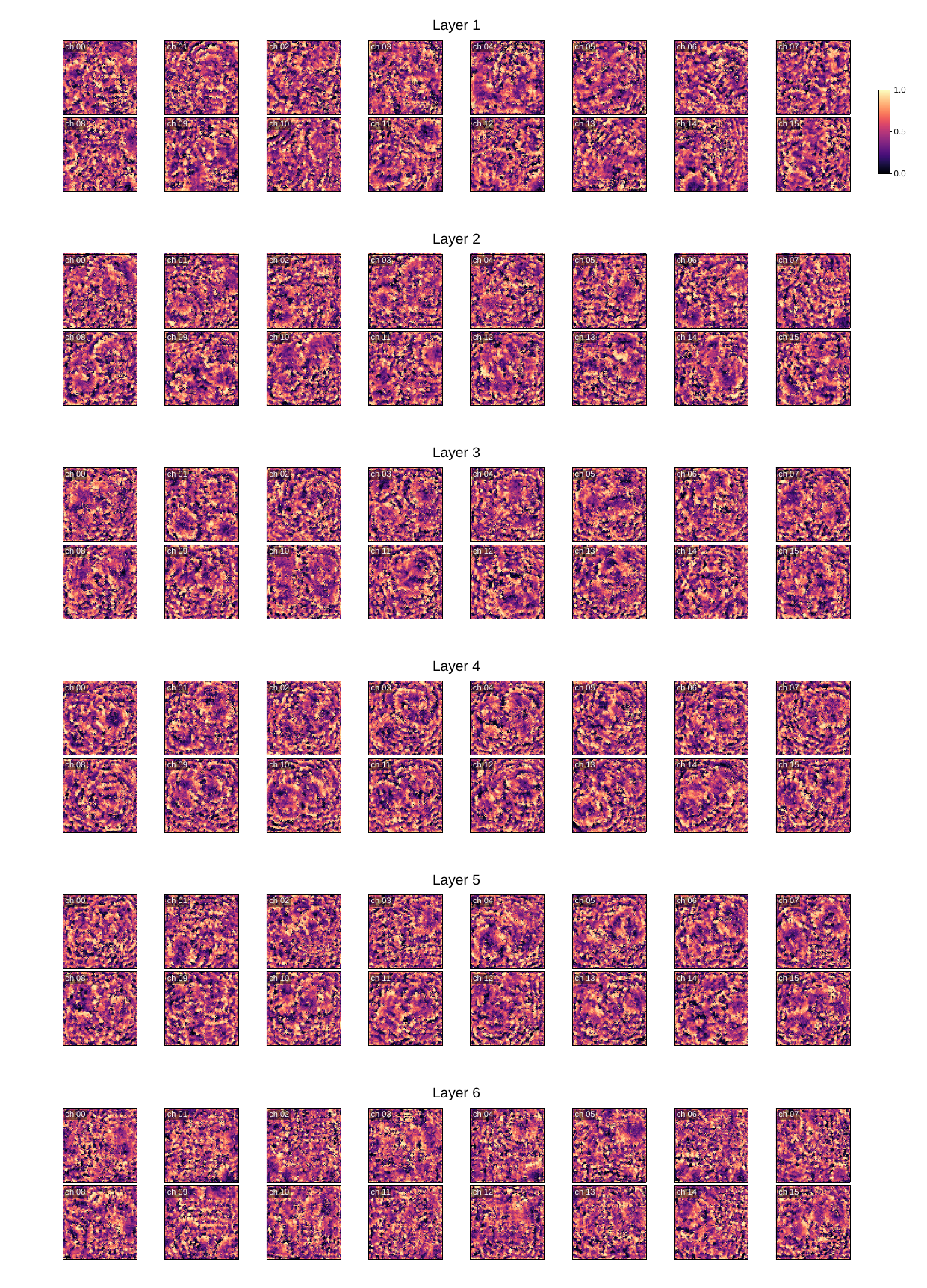}
\caption{\textbf{Trained phase masks for vision-language captioning.} The optimized phase masks are shown for the six optical layers and all 16 spatial channels.}
\label{fig:phase_masks_vlm}
\end{figure}

Figures~\ref{fig:phase_masks_independent_mnist}--\ref{fig:phase_masks_vlm} show the trained phase masks obtained for the five optical vision tasks summarized in Tables~\ref{tab:task_interface} and~\ref{tab:optimization_protocol}. For each optical layer, the 16 channel-specific masks are arranged from ch~00 to ch~15. The resulting spatial phase distributions are task dependent because the masks are optimized jointly with the corresponding optical readout, channel-mixing modules and task objective. In all figures, color encodes the normalized phase command applied to the spatial light modulator.

\clearpage

\clearpage

\section{Results comparison with state-of-the-art}
\label{sec:supp_results_comparison}

Table~\ref{tab:state_of_the_art_comparison} compares the present model with representative experimental optical and optoelectronic neural networks reported in recent high-impact journals. Because these systems differ in their optical--electronic partition, trainable-parameter convention and evaluation protocol, the table reports optical and electronic parameters separately and summarizes the principal architectural differences. The comparison is intended as a system-level contextual benchmark rather than as a strictly controlled accuracy ranking.

\begin{landscape}
\begin{center}
\scriptsize
\setlength{\tabcolsep}{2pt}
\renewcommand{\arraystretch}{1.18}
\begin{longtable}{>{\raggedright\arraybackslash}p{0.14\linewidth} >{\centering\arraybackslash}p{0.09\linewidth} >{\centering\arraybackslash}p{0.09\linewidth} >{\centering\arraybackslash}p{0.11\linewidth} >{\centering\arraybackslash}p{0.12\linewidth} >{\centering\arraybackslash}p{0.12\linewidth} >{\raggedright\arraybackslash}p{0.25\linewidth}}
\caption{\textbf{Comparison with representative experimental optical and optoelectronic neural networks.} Reported classification values are experimental accuracies from the cited studies, while the results of this work are the best held-out validation values from the corresponding saved runs. Optical and electronic parameter counts refer to task-trainable model parameters; fixed optical hardware, fixed codebooks and the training-only surrogate are excluded.}
\label{tab:state_of_the_art_comparison}\\
\toprule
\textbf{System} & \textbf{MNIST} & \textbf{Fashion-MNIST} & \textbf{Facial-keypoint regression} & \textbf{Trainable optical params} & \textbf{Trainable electronic params} & \textbf{Implementation details} \\
\midrule
\endfirsthead
\toprule
\textbf{System} & \textbf{MNIST} & \textbf{Fashion-MNIST} & \textbf{Facial-keypoint regression} & \textbf{Trainable optical params} & \textbf{Trainable electronic params} & \textbf{Implementation details} \\
\midrule
\endhead
Zhu \textit{et al.}\ \cite{zhu2022space} & 91.4\% & 80.2\% & -- & 10 complex weights (20 real DOF) & 160 complex weights (320 real DOF) & A Fourier-domain crop produces 16 complex features. A learned digital $16\!\rightarrow\!10$ complex transform precedes a 10-coefficient complex circulant optical transformation. Values are the detailed Fig.~5 experimental results. \\
\midrule
Yildirim \textit{et al.}\ \cite{yildirim2024nonlinear} & $87.90\pm0.06$\% & $83.84\pm0.24$\% & -- & 45,000 & Not exactly reported; 160--170 derived for Fashion-MNIST & Four $75\times75$ optical planes use a trainable scale and bias at every site. The pooled optical intensity is passed to a co-trained single-layer digital classifier. \\
\midrule
Xue \textit{et al.}\ \cite{xue2024fully} & -- & 92.5\% & -- & 1,280,000 & 0 & Eight sequentially emulated linear optical layers, each using a $400\times400$ pure-phase trainable mask, followed by direct optical output-region readout without a learned electronic classifier. \\
\midrule
Liang \textit{et al.}\ \cite{liang2026highclockrate} & 93.75\% & 80.75\% & -- & 81 & 9,000 & Nine bias-free $3\times3$ optical convolution kernels are followed by parameter-free digital ReLU and a bias-free digital $900\!\rightarrow\!10$ classifier. \\
\midrule
Xia \textit{et al.}\ \cite{xia2024nonlinear} & -- & -- & 1.06-pixel mean error & 0 task-trained & $\approx310{,}000$ & A fixed passive recurrent-scattering optical encoder produces 16 measured speckle modes, which are decoded by a nine-layer digital multilayer perceptron. \\
\midrule
Digital CNN baseline in Ref.~\cite{xia2024nonlinear} & -- & -- & $\approx1.23$-pixel mean error & -- & $>74$ million & Two convolutional layers, one pooling layer and three fully connected layers, requiring approximately 83 million operations. \\
\midrule
\textbf{This work} & \textbf{91.50\%} & \textbf{83.17\%} & \textbf{2.461-pixel coordinate RMSE} & \textbf{1,003,520 for each task} & \textbf{MNIST: 16; Fashion-MNIST: 10,112; Facial: 184,080} & MNIST uses optical YES/NO measurements and fixed class-code decoding. Fashion-MNIST uses electronic input adaptation and inter-layer mixing. Facial-keypoint regression uses electronic spatial mixers and linear coordinate heads. \\
\bottomrule
\end{longtable}

\begin{minipage}{0.97\linewidth}
\footnotesize
\textit{Notes.} A complex parameter in Zhu \textit{et al.} contains two real-valued degrees of freedom. The electronic count for Yildirim \textit{et al.} is not stated explicitly: the reported $4\times4$ pooling followed by a $16\!\rightarrow\!10$ linear classifier implies 160 weights, or 170 parameters if ten biases are included. The Xue and Liang parameter totals are derived directly from their reported architectures. Xia \textit{et al.} report approximately 4.1 million addressable DMD mirrors, but these encode the input and are not task-trained optical weights; only the digital decoder is optimized. Their ``mean pixel error'' is retained as the authors' metric because its aggregation is not defined in sufficient detail to equate it with the coordinate RMSE used here. For this work, the optical count follows the phase-parameter convention used throughout the manuscript and reports the five $16\times112\times112$ trainable phase arrays. The class-code lookup table and the differentiable surrogate used only during training are not included in the inference parameter count.
\end{minipage}
\end{center}
\end{landscape}

\clearpage

\section{Energy consumption and TOPS/W}
\label{sec:supp_energy_efficiency}

Table~\ref{tab:power_consumption} summarizes an estimate of the typical consumption of each device.
CW HeNe lasers require a DC current of 3 to 20 mA \cite{Sakurai1972}, the Holoeye devices consumption are given by the power supply output power (Holoeye Eris) and the maximum rating of USB2.0 ports (Holoeye HEO). Finally, the Basler power consumption is provided by the manufacturer specification.

\begin{center}
\scriptsize
\setlength{\tabcolsep}{3pt}
\renewcommand{\arraystretch}{1.25}
\begin{longtable}{>{\centering\arraybackslash}p{0.1\textwidth} >{\centering\arraybackslash}p{0.15\textwidth} >{\centering\arraybackslash}p{0.15\textwidth} >{\centering\arraybackslash}p{0.15\textwidth} >{\centering\arraybackslash}p{0.18\textwidth}}
\caption{\textbf{Power consumption estimate}. This table list the continuous power consumption drawn by each device in our experiment.}
\label{tab:power_consumption}\\
\toprule
HeNe Laser &  Holoeye HEO & Holoeye ERIS & Basler a2A1920 & \textbf{Total} \\
\midrule
5W & 2.5W & 20W & 2.5W  & \textbf{30W}\\

\bottomrule
\end{longtable}
\end{center}

We now consider the equivalent complex MAC operations performed by the optical propagation based on Eq.~\ref{eq:physical_channel_forward}:

Let \(n_{\mathrm{in}}=N_{\mathrm{in},x}N_{\mathrm{in},y}\) be the total number of input pixels from the microdisplay, and let
\(n_{\mathrm{out}}=N_{\mathrm{out},x}N_{\mathrm{out},y}\) be the total number of output pixels recorded by the camera.
Under the convention that one real multiplication or one real addition counts as one operation, one complex multiplication requires 4 real multiplications + 2 real additions = 6 operations.
Accumulating the complex products requires two additional real additions per subsequent term for a total of 8 operations.

Let \(T_{h_1}\in\mathbb{C}^{n_{\mathrm{in}}\times n_{\mathrm{in}}}\) and \(T_{\Delta z}\in\mathbb{C}^{n_{\mathrm{out}}\times n_{\mathrm{in}}}\) be the block-Toeplitz propagation matrices associated with \(h_1\) and \(h_{\Delta z}\), respectively.
Define \(M_{h_1}\) and \(M_{\Delta z}\) as the numbers of nonzero elements in \(T_{h_1}\) and \(T_{\Delta z}\), respectively.
The total number of operations for both convolutions is \(N_{\mathrm{prop}}=8(M_{h_1}+M_{\Delta z})-2(n_{\mathrm{in}}+n_{\mathrm{out}})=8M_{\Delta z}-2n_{\mathrm{out}}+2n_{\mathrm{in}}\).
Moreover, we take into account the structural-nonlinearity encoding of the weights \(\theta_{\mathrm{phys},c}\), which contributes \(N_{\mathrm{struct}}=2n_{\mathrm{in}}\) operations.
Before the final propagation with \(T_{\Delta z}\), the elementwise operation contributes \(N_{\odot}=6n_{\mathrm{in}}\) operations.
The camera measures the intensity, performing the operation:

\begin{equation*}
I_j=|E_j|^2=\operatorname{Re}(E_j)^2+\operatorname{Im}(E_j)^2.
\end{equation*}

This readout nonlinearity requires two real multiplications and one real addition per output pixel, so it contributes \(3n_{\mathrm{out}}\) operations. The \(3n_{\mathrm{out}}\) term is not a fitted nonlinear-activation cost; it is the real-operation-equivalent cost of the camera square-law intensity measurement at the output plane.
Therefore, the per-channel, per-pass operation count is:

\begin{equation*}
N_{\mathrm{pass},\mathrm{ch}}
=2n_{\mathrm{in}}+2n_{\mathrm{in}}+6n_{\mathrm{in}}+
\left(8M_{\Delta z}-2n_{\mathrm{out}}\right)+3n_{\mathrm{out}}
=10n_{\mathrm{in}}+8M_{\Delta z}+n_{\mathrm{out}}.
\end{equation*}

For the current hardware estimate, the input field size per channel is \(112\times112\) pixels, while the output camera readout size per channel is \(168\times168\) pixels.

Hence, \(n_{\mathrm{in}}=N_{\mathrm{in},x}N_{\mathrm{in},y}=112\times112=12{,}544\) and \(n_{\mathrm{out}}=N_{\mathrm{out},x}N_{\mathrm{out},y}=168\times168=28{,}224\).
The effective kernel side length is \(K_{\mathrm{side}}=N_{\mathrm{out}}-N_{\mathrm{in}}+1=57\).
The PSF \(h_{\Delta z}\) contains \(K=57\times57=3{,}249\) complex coefficients.
For full convolution, every active input pixel contributes through all \(K\) kernel coefficients.
Therefore, the total number of nonzero entries in the block-Toeplitz operator is \(M_{\Delta z}=n_{\mathrm{in}}K=40{,}755{,}456\).
After substitution, we obtain \(N_{\mathrm{pass},\mathrm{ch}}=3.262\times10^{8}\) operations per channel per optical pass.
There are \(N_{\mathrm{c}}=16\) simultaneous spatial channels, giving \(N_{\mathrm{pass},\mathrm{total}}=5.219\times10^{9}\) operations per optical pass.
At an effective forward-pass rate of \(f_{\mathrm{eff}}=\SI{60}{\hertz}\), the operation rate and energy efficiency are therefore
\begin{align*}
    \mathrm{TOPS} &= N_{\mathrm{pass},\mathrm{total}}f_{\mathrm{eff}}\times10^{-12}=0.313, \\
    \mathrm{TOPS/W} &= \frac{0.313}{30}=0.0104.
\end{align*}

The current system uses only \(N_{\mathrm{c}}=16\) channels; this number can be increased by exploiting the full degrees of freedom of the SLMs.
While keeping \(N_{\mathrm{in}}=112\), we can decrease the anti-crosstalk padding from 100 pixels to 57 pixels.
The smallest SLM has \(1080\times1920\) pixels and can fit 6 images of size \(N_{\mathrm{in},y}=112\) pixels along its height, with 5 gaps of 57 pixels occupying a total of 957 pixels.
Similarly, along the width, the screen may fit 14 images of size \(N_{\mathrm{in},x}=112\) pixels.
This would increase the total number of instantaneous channels to \(N_{\mathrm{c}}=84\).
Therefore, the corresponding operation rate would be \(\mathrm{TOPS}=1.64\).

Moreover, our current SLMs limit the effective forward-pass rate to \(f_{\mathrm{eff}}=\SI{60}{\hertz}\).
Recent progress in these devices can increase this rate to \(f_{\mathrm{eff}}=\SIrange{5}{50}{\kilo\hertz}\) \cite{Griffin2025, Bartlett2019, Rocha2025}.
This results in theoretical values of \(\mathrm{TOPS}=1369\) and \(\mathrm{TOPS/W}=45.57\), assuming a similar order of power consumption.

\begingroup
\scriptsize
\setlength{\tabcolsep}{5pt}
\renewcommand{\arraystretch}{1.15}
\begin{longtable}{>{\raggedright\arraybackslash}p{0.42\textwidth} >{\centering\arraybackslash}p{0.16\textwidth} >{\centering\arraybackslash}p{0.16\textwidth}}
\caption{\textbf{Power-consumption and computational-efficiency comparison.} Reported TOPS and TOPS/W values for representative computing platforms and for the current and projected configurations of our system. Rows corresponding to this work are shaded in gray.}
\label{tab:comp_power_consumption}\\
\toprule
\textbf{Platform or configuration} & \textbf{TOPS} & \textbf{TOPS/W} \\
\midrule
Dong et al. (2023) \cite{dong2023higher} & 0.014 & 0.089\\
Xu et al. (2024) \cite{xu2024taichi} & -- & 160\\
Hua et al. (2025) \cite{hua2025integrated} & 8.19 & 2.38\\
Ahmed et al. (2025) \cite{ahmed2025universal} & 65.5 & 0.818\\
NVIDIA A100 80GB SXM \href{https://www.nvidia.com/en-us/data-center/a100/}{[FP64 specification]} & 19.5 & 0.049\\
NVIDIA GB200 Grace Blackwell \href{https://www.nvidia.com/en-us/data-center/gb200-nvl72/}{[FP64 specification]} & 80 & 0.066\\
\midrule
\rowcolor{gray!15}\textbf{Current system} & \textbf{0.313} & \textbf{0.0104}\\
\rowcolor{gray!15}\textbf{Optimized multi-channel} & \textbf{1.64} & \textbf{0.054}\\
\rowcolor{gray!15}\textbf{Faster SLM} & \textbf{1369} & \textbf{45.57}\\
\bottomrule
\end{longtable}
\endgroup

\end{document}